\documentclass[journal]{IEEEtran}
\usepackage{times}
\usepackage{epsfig}
\usepackage{graphicx}
\usepackage{graphicx}
\usepackage{amsmath}
\usepackage{amssymb}
\usepackage{multirow}
\usepackage{bbding}
\usepackage{color}
\usepackage{stfloats}
\usepackage{floatrow}
\usepackage{algorithm}
\usepackage{algorithmic}
\usepackage{adjustbox}
\usepackage{bm}
\usepackage{booktabs}
\usepackage{threeparttable}
\usepackage{lettrine}
\usepackage[switch]{lineno} 
\usepackage[numbers,sort&compress]{natbib}

\begin{document}

\title{JAS-GAN: Generative Adversarial Network Based Joint Atrium and Scar Segmentations on Unbalanced Atrial Targets}

\author{ Jun Chen, Guang Yang,  Habib Khan, Heye Zhang, Yanping Zhang, Shu Zhao, Raad Mohiaddin, Tom Wong, David Firmin, Jennifer Keegan.
\thanks{This study was supported in part by the Key-Area Research and Development Program of Guangdong Province (2019B010110001), in part by the National Natural Science Foundation of China (61771464, U1801265), in part by the Key Program for International Cooperation Projects of Guangdong Province (2018A050506031), in part by the Guangdong Natural Science Funds for Distinguished Young Scholar (2019B151502031), in part by the British Heart Foundation (Project Number: TG/18/5/34111, PG/16/78/32402), in part by the Innovative Medicines Initiative (H2020-JTI-IMI2 101005122), and in part by the ERC H2020 (H2020-SC1-FA-DTS-2019-1 952172).}
\thanks{J. Chen and G. Yang are co-first authors contributed equally to this work. H. Zhang and Y. Zhang are corresponding authors.}
\thanks{J. Chen and H. Zhang are with the School of Biomedical Engineering, Sun Yat-sen University, Guangzhou, China (e-mail: chenj657@mail2.sysu.edu.cn; zhangheye@mail.sysu.edu.cn).}
\thanks{G. Yang, Raad Mohiaddin, Tom Wong, David Firmin and Jennifer Keegan are with the Cardiovascular Research Centre, Royal Brompton Hospital, SW3 6NP, London, U.K and the National Heart and Lung Institute, Imperial College London, London, SW7 2AZ, U.K. (e-mail: g.yang@imperial.ac.uk; R.Mohiaddin@rbht.nhs.uk; T.Wong2@rbht.nhs.uk; D.firmin@imperial.ac.uk; J.keegan@imperial.ac.uk).}
\thanks{H. Khan is with the Cardiovascular Research Centre, Royal Brompton Hospital, SW3 6NP, London, U.K. (e-mail: hurkhan@msn.com).}
\thanks{Y. Zhang and S. Zhao are with the School of Computer Science and Technology, Anhui University, Hefei, China (e-mail: zhangyp2@gmail.com; zhaoshuzs2002@hotmail.com).}
}

\maketitle
\begin{abstract}
Automated and accurate segmentations of left atrium (LA) and atrial scars from late gadolinium-enhanced cardiac magnetic resonance (LGE CMR) images are in high demand for quantifying atrial scars. The previous quantification of atrial scars relies on a two-phase segmentation for LA and atrial scars due to their large volume difference (unbalanced atrial targets). In this paper, we propose an inter-cascade generative adversarial network, namely JAS-GAN, to segment the unbalanced atrial targets from LGE CMR images automatically and accurately in an end-to-end way. Firstly, JAS-GAN investigates an adaptive attention cascade to automatically correlate the segmentation tasks of the unbalanced atrial targets. The adaptive attention cascade mainly models the inclusion relationship of the two unbalanced atrial targets, where the estimated LA acts as the attention map to adaptively focus on the small atrial scars roughly. Then, an adversarial regularization is applied to the segmentation tasks of the unbalanced atrial targets for making a consistent optimization. It mainly forces the estimated joint distribution of LA and atrial scars to match the real ones. We evaluated the performance of our JAS-GAN on a 3D LGE CMR dataset with 192 scans. Compared with the state-of-the-art methods, our proposed approach yielded better segmentation performance (Average Dice Similarity Coefficient (DSC) values of 0.946 and 0.821 for LA and atrial scars, respectively), which indicated the effectiveness of our proposed approach for segmenting unbalanced atrial targets.
\end{abstract}


\IEEEpeerreviewmaketitle

\section{Introduction}
\IEEEPARstart{A}{utomated} and accurate segmentations of left atrium (LA) and atrial scars from late gadolinium enhanced cardiac magnetic resonance (LGE CMR) are crucial for the quantification of atrial scars. The quantification of atrial scars usually requires the segmentations of the LA and atrial scars to obtain an accurate estimation of the scar percentage \cite{ravanelli2014novel}, helping the treatment stratification of patients with atrial fibrillation (AF) before and after radio-frequency catheter ablation \cite{Karim2013Evaluation,vergara2011tailored}. Clinically, LGE CMR imaging allows the visualization of scar tissues through the amount of contrast agent left due to differences in interstitial cell structures \cite{siebermair2017assessment}. Thus, LGE CMR has emerged as a promising technique to non-invasively detect and locate atrial scars to further provide the accurate quantification of atrial scars \cite{siebermair2017assessment}. In clinical practice, this generally relies on manual segmentations of both the LA and atrial scars \cite{khurram2016left}, which is time-consuming. Automated segmentations of the LA and atrial scars from LGE CMR images would facilitate the rapid and reproducible quantification of atrial scars.

\begin{figure}[!hbtp]
\begin{center}
   \includegraphics[width=1\linewidth]{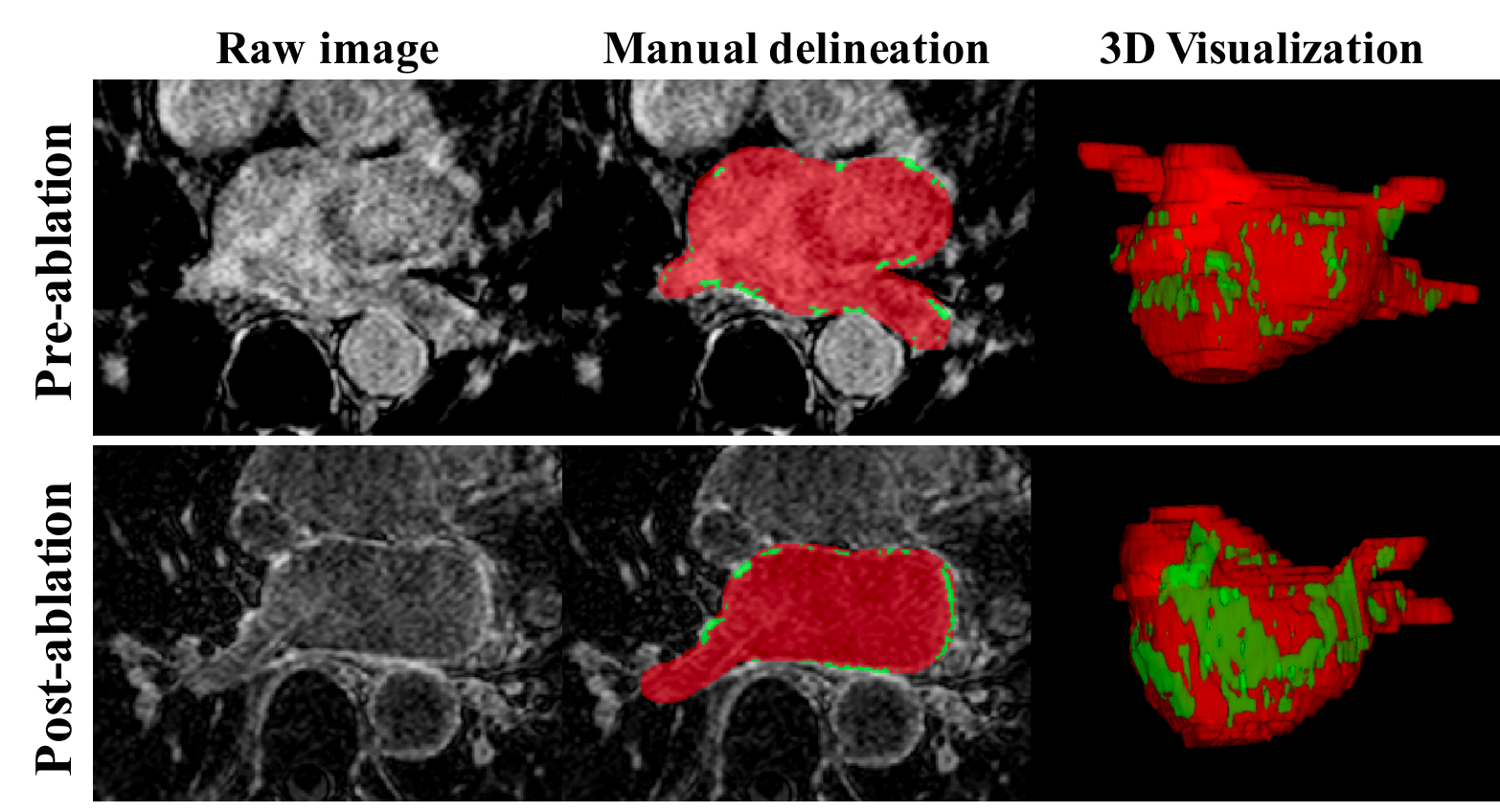}
\end{center}
\caption{Examples of transverse LGE CMR slices (left) together with manual segmentations (middle) and 3D visualization (right) in a pre-ablation scan (top raw) and a post-ablation scan (bottom raw) from two patients. The red regions denote the LA while the green ones denote the atrial scars.}
\label{fig:im_vs} 
\end{figure}

However, automated and accurate segmentations of LA and atrial scars from LGE CMR images are two very challenging tasks due to the complexities of the two unbalanced targets with significant volume contrast as shown in Fig. \ref{fig:im_vs}. Firstly, for the segmentation task of LA, the LGE CMR imaging technology is generally used to visualize scar tissue by enhancing its signal intensity. This gives rise to the attenuated contrast in non-diseased tissue \cite{xiong2020global}. The attenuated contrast in healthy LA reduces the visibility of the LA boundaries, which limits the usage of edge and region based methods for the automated and accurate segmentation of the LA. Secondly, for the segmentation task of atrial scars, atrial scars occupy only a very small portion of LA volume. They are therefore highly susceptible to noise interference. Besides, compared with the voxels in the background, the amount of information available on the small atrial scars is very limited, which results in severe class-imbalance problems for hindering the automated and accurate segmentation of atrial scars. Furthermore, there are many other nearby tissues (aortic wall, oesophagus and other tissues) that are enhanced by LGE CMR imaging along with atrial scars, which also can interfere with the accurate recognition of the atrial scars. To tackle these difficulties, most of the work done in this field has focused on a separated two-phase segmentation framework, where the LA is obtained first followed by the delineation of the small atrial scars. This two-phase segmentation framework is limited to the inefficiency and error accumulation problem.

In order to overcome the issues mentioned above, we investigate end-to-end joint learning for two semantic segmentation tasks of the unbalanced targets of LA and atrial scars. Because the atrial scars are located in the LA wall, there exists an inclusion relationship between the large LA and small atrial scars. We can make full use of the inclusion relationship to mine the dependent correlation between the segmentation tasks of LA and atrial scars for their joint learning. However, the LA and atrial scars are unbalanced targets  as shown in Fig. \ref{fig:im_vs}, which can bring the problem of inconsistent target learning. Hence, we further investigate the adversarial learning for consistent target learning  \cite{isola2017image,hung2019adversarial}. 

In this paper, we propose a \textbf{J}oint \textbf{A}trium (i.e., LA) and \textbf{S}car (i.e., atrial scars) segmentation framework based on an inter-cascade \textbf{G}enerative \textbf{A}dversarial \textbf{N}etwork, namely \textbf{JAS-GAN}, from LGE CMR images. In our proposed JAS-GAN, the inclusion relationship between the unbalanced targets of large LA and small atrial scars is effectively mined for their accurate and joint segmentations. Our proposed JAS-GAN consists of an adaptive attention cascade network and a joint discriminative network: (1) The adaptive attention cascade network contains an encoder-decoder module for LA segmentation and a residual network for atrial scars segmentation. The two modules are cascaded through an adaptive attention connection to model the spatial correlation of the LA and atrial scars. The adaptive attention connection makes full use of the segmented LA as an attention map to further roughly focus on the small atrial scars in an end-to-end way. (2) The joint discriminative network further transforms the segmentation problem of pixel-level classification for the unbalanced targets of LA and atrial scars into a problem of pixel-level identification, that is, whether the pixels at the same position in the LA and atrial scar segmentation maps are produced by the adaptive attention cascade network or from the ground truth label maps. It mainly employs an adversarial regularization to force the estimated joint distribution of LA and atrial scars to match the real ones,  which can provide a consistent optimization for the segmentation task learning of unbalanced atrial targets.

Finally, the contributions of our framework can be summarized as follows:
\begin{enumerate}
\item We propose an end-to-end segmentation framework for the LA and atrial scars to facilitate the rapid and reproducible quantification of atrial scars. The framework can further provide the essential guidance for clinicians to analyze the structures of LA and atrial scars directly from 3D LGE CMR images.

\item We propose an inter-cascade adversarial learning paradigm to mine the relationship of unbalanced targets automatically by modelling their position and joint distribution.

\item We have conducted comprehensive experiments on a 3D LGE CMR dataset with 192 scans for validating our proposed JAS-GAN. The results demonstrated the better performance of JAS-GAN over the state-of-the-art and traditional methods, which indicated the feasibility of unbalanced atrial targets segmentation framework.
\end{enumerate}

\section{Related Work}
\subsection{Two-Phase Segmentation Methods for Quantifying Atrial Scars}
Currently, the most related methods to the quantification of atrial scars rely on a two-phase sequential segmentation of the LA and atrial scars \cite{Karim2013Evaluation,Yang2018Fully}. These methods are inadequate to achieve accurate quantification as the segmentations of the LA and atrial scars are handled separately. There is no feedback loop existing between them during model learning, thus leading to the error accumulation problem. 

In these methods, segmenting the LA cavity or LA wall is usually the first step to further locate the atrial scars. Furthermore, instead of directly segmenting the LA cavity or LA wall from LGE CMR scans, some methods rely on a separately acquired breath-hold magnetic resonance angiogram (MRA) study or on a respiratory and cardiac gated 3D Roadmap acquisition for LA segmentation. Then, they registered the segmented LA to the LGE CMR acquisition for the delineation of atrial scars. Previously proposed methods for LA wall segmentation include (1) manual segmentation \cite{Karim2013Evaluation,Perry2015Automatic,ravanelli2014novel}, which is tedious and inefficient, (2) segmentation of the LA cavity followed by some morphological dilations for LA wall extraction \cite{karim2014method}, and (3) automated or semi-automatic LA wall segmentation, e.g., active contour based segmentation \cite{Karim2013Evaluation}. Furthermore, many automated methods have been proposed for segmenting the LA \cite{Mortazi2017CardiacNET,Tobon2015Benchmark,XiongFully,chen2019discriminative,yu2019uncertainty,zhuang2019evaluation}. However, they have not yet been further applied to the quantification of the atrial scars. 

Based on the segmented LA wall, histogram analysis, thresholding, k-means clustering, and graph-cuts based unsupervised methods have been applied to segment atrial scars \cite{Karim2013Evaluation}. However, these unsupervised learning methods are susceptible to various image quality and noise conditions. Yang et al. \cite{Yang2018Fully} proposed deep learning and support vector machines based supervised classification methods to segment atrial scars and achieved better results. However, it still relies on a two-phase segmentation for LA and atrial scars.

\begin{figure*}[!ht]
	\begin{center}
	 \includegraphics[width=1\textwidth]{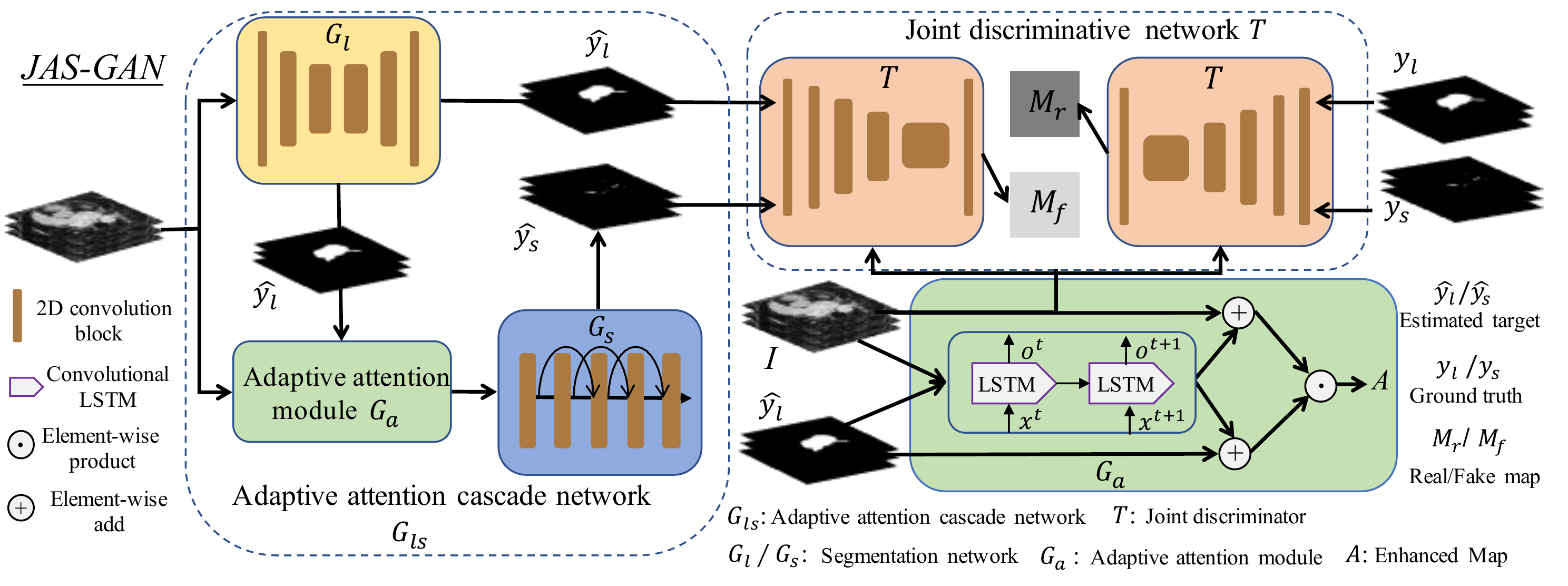}
	\end{center}
	\caption{The architecture of our proposed JAS-GAN for the joint segmentations of unbalanced atrial targets. Each 3D input volume is sliced at axial plane then fed into the JAS-GAN. An adaptive attention cascade network correlates the spatial location of LA and small atrial scars. A joint discriminative network is designed to regularize the adaptive attention cascade network to produce matched joint distribution of unbalanced atrial targets.}
	\label{fig:JAS-GAN}
\end{figure*}

\subsection{Cascade Learning}
Cascade is an efficient structure to improve performance for the deep learning based single task or multiple tasks solver \cite{li2017not,murthy2016deep,cai2018cascade,dai2016instance,ouyang2017chained,lin2017cascaded,chen2019hybrid}, which has been widely used in various applications including classification, detection and segmentation. For single task learning, the cascade can be divided into multi-stage learning. The latter stages can focus on more accurate learning to improve the performance stage by stage and achieve a faster inference. For the multiple tasks problem, the tasks are designed in a cascade manner that the tasks at a later stage depending on the output of an earlier stage.

\section{Methodology}
\textbf{Goal:} Learn unbalanced atrial targets segmentation model by using LGE CMR images.

\textbf{Notation:} $I\in\Re^{H\times W\times C}$ represents the input image with size of $H\times W$ and channel of $C$. The subscripts of $l$ and $s$ in notations denote the LA and the atrial scars respectively. $\widehat{y}_{l}$ and $\widehat{y}_{s}$ represent the estimated LA and the estimated atrial scars respectively, while $\widehat{Y}_{ls}=(\widehat{y}_{l},\widehat{y}_{s})$ represents the both estimated LA and atrial scars. $y_{l}$ and $y_{s}$ represent the ground truth of LA and atrial scars respectively, while $Y_{ls}=(y_{l},y_{s})$ represents the ground truth of both LA and atrial scars. $F_{w1}$ and $F_{w2}$ denote the weight maps. A represents the enhanced map. $M_{r}\in[0,1]^{H\times W\times 1}$ and $M_{f}\in[0,1]^{H\times W\times 1}$ denote the real confidence map and the fake confidence map with both size of $H\times W$ and channel of 1. $G_{l}$, $G_{s}$ and $G_{a}$ represents the encoder-decoder network (EDN), the residual network (RN) and the convolutional long short-term memory (convLSTM) based adaptive attention connection module (AC), respectively. $G_{ls}$ represents the combination of $G_{l}$, $G_{a}$ and $G_{s}$. $T$ represents the joint discriminative network.

\subsection{An Overview of JAS-GAN}
Fig. \ref{fig:JAS-GAN} displays our proposed JAS-GAN. JAS-GAN mainly comprises of an adaptive attention cascade network $G_{ls}$ and a joint discriminative network $T$. Specifically, the adaptive attention cascade network $G_{ls}$ is designed for the joint segmentations of LA and atrial scars. The $G_{ls}$ consists of an encoder-decoder network $G_{l}$ for LA segmentation, a residual network $G_{s}$ for atrial scars segmentation and an adaptive attention connection module $G_{a}$ for constructing cascade connection that the $G_{l}$ and the $G_{s}$ are cascaded by the $G_{a}$. The joint discriminative network $T$ conditioned on the image is designed to force the $G_{ls}$ to produce a correct joint distribution of LA and atrial scars.

\subsection{Adaptive Attention Cascade Network for Unbalanced Atrial Targets Simultaneous Estimation}
We model the inclusion relationship of LA and atrial scars to build a cascade segmentation network for their joint segmentation. Firstly, consider an encoder-decoder establishes a mapping $G_{l}:I\rightarrow \widehat{y}_{l}$ to estimate $\widehat{y}_{l}$ from $I$ directly. We notice that small atrial scars are located in the LA wall, the LA can be taken as  prior  knowledge  to  constrain  the learnable  area  of  atrial  scars,  which reduces  the interference of external noise of LA for atrial scar identification. Therefore, we model their inclusion relationship to leverage the LA to focus on the small atrial scars roughly. Because the voxel values of predicted $\widehat{y}_{l}$ range from $[0,1]$, the $\widehat{y}_{l}$ can be used as an attention map that 1 represents the full attention while the 0 denotes no attention to pay attention to atrial scars roughly on $I$. However, atrial scars distributed beside the border of the LA as shown in Fig. \ref{fig:im_vs}. If the $G_{l}$ produces an under-segmented LA, the general attention operation may weaken the atrial scars partially or completely. Therefore, we further investigate an adaptive attention module $G_{a}:(I,\widehat{y}_{l})\rightarrow A$ to estimate an enhanced map A for scar identification by mining the relationship of $\widehat{y}_{l}$ and corresponding $I$ for adaptively adjusting the attention operation. In detail, $G_{a}$ firstly establishes a mapping $(I,\widehat{y}_{l})\rightarrow (F_{w1},F_{w2})$ based on a convLSTM to learn the relationship of $I$ and $\widehat{y}_{l}$ and estimate two weight maps of $F_{w1}$ and $F_{w2}$. The two weight maps of $F_{w1}$ and $F_{w2}$ are then used to adaptively adjust the attention operation:
\begin{equation}
\begin{split}
A=&(F_{w1}+I)\cdot(F_{w2}+\widehat{y}_{l})\\
=&I\cdot \widehat{y}_{l} + I\cdot F_{w2} +\widehat{y}_{l}\cdot F_{w1} +F_{w1}\cdot F_{w2}
\end{split}
\end{equation}
where the $\cdot$ denotes the element-wise product. As shown in Equation (1), the terms of $I\cdot F_{w2} +\widehat{y}_{l}\cdot F_{w1} +F_{w1}\cdot F_{w2}$ adaptively adjust the  general attention operation of $I\cdot \widehat{y}_{l}$. Based on the obtained $A$, we can separate atrial scars from the surrounding enhanced tissues and organs with highly similar intensities to scars. Then we consider a residual network to establish an another mapping $G_{s}:A\rightarrow \widehat{y}_{s}$ to estimate $\widehat{y}_{s}$. The residual network discards the downsampling operation to avoid information loss of small atrial targets.

Therefore, to achieve the joint estimation for $\widehat{y}_{l}$ and $\widehat{y}_{s}$, we directly concatenate $G_{l}$ and $G_{s}$ by establishing a function: $G_{ls}:I\rightarrow \widehat{Y}_{ls}$, defined by $G_{ls}=G_{s}\circ G_{a}\circ G_{l}$. In this case, $G_{l}$ firstly estimates the $\widehat{y}_{l}$ from input image $I$. Then $G_{a}$ produces an adaptive attention map $A$ from the estimated $\widehat{y}_{l}$ and the input image $I$. Finally, $G_{s}$ estimates the $\widehat{y}_{s}$ from $A$. The adaptive attention cascade network $G_{ls}$ integrates the segmentations of LA and atrial scars into one step by the seamless cascade connection. Such connection results in an optimal model learning to leverage the large LA to catch the small atrial scars. It further can automatically relieve the error accumulation and noisy interference for the accurate segmentation of small atrial scars.

\subsection{Joint Discriminative Network for Adversarial Regularization}
The adaptive attention cascade network $G_{ls}$ is optimized to produce the right class label at each voxel location independently. We further investigate an adversarial learning to transform the unbalanced target segmentation that classifies large LA and small atrial scars to identify whether the pixels at the same position in the LA and atrial scar segmentation maps are produced by the $G_{ls}$ or from the ground truth. The adversarial learning regularizes the adaptive attention cascade network to force the estimated joint distribution of LA and atrial scars to match the real ones. Specifically, consider a joint discriminative network $T$ conditioned on $I$ establishes a mapping $T:(Y_{ls},I)\rightarrow M_{r}$ and $T:(\widehat{Y}_{ls},I)\rightarrow M_{f}$. Each pixel ($M^{(i,j)}_{r}$ and $M^{(i,j)}_{f}$, where $(i,j)$ denotes the spatial position of the map with $i\in [0,H]$ and $j\in [0,W]$) of the confidence map represents whether that the pixels at the same position in the LA and atrial scar segmentation maps are sampled from the ground truth label or produced by the $G_{ls}$. The prior distributions on the two atrial targets and image are denoted as the $p(Y_{ls})$ and $p(I)$, respectively. Then, we consider the following objectives:
\begin{equation}
\begin{split}
\mathop{\min_{G_{ls}}\max_{T}}\ &E_{Y_{ls}\sim p(Y_{ls})}\sum_{i,j}[log\ \sigma(T(Y_{ls},I)^{(i,j)})]\\
&+E_{I\sim p(I)}\sum_{i,j}[1-log\ \sigma(T(G_{ls}(I),I)^{(i,j)})]
\end{split}
\end{equation}
where $\sigma$ denotes the sigmoid function. $T(Y_{ls},I)^{(i,j)}$ is the confidence map of $M_{r}$ at location $(i,j)$ while $T(G_{ls}(I),I)^{(i,j)}$ is the confidence map of $M_{f}$ at location $(i,j)$. The Equation (2) makes the learning of $G_{ls}$ and $T$ a dynamic adversarial process to regularize the $G_{ls}$ to produce the estimated  joint  distribution  of  LA  and  atrial  scars  to  match  the  real  ones. Specifically, on the one hand, the $T$ tries to distinguish the estimated LA and atrial scars from the ground truth by 
\begin{equation}
\begin{split}
\mathop{\max_{T}}\ L_{d}&=E_{Y_{ls}\sim p(Y_{ls})}\sum_{i,j}[log\ \sigma(T(Y_{ls},I)^{(i,j)})]\\
&+E_{I\sim p(I)}\sum_{i,j}[1-log\ \sigma(T(G_{ls}(I),I)^{(i,j)})]
\end{split}
\end{equation}
On the other hand, the mismatches between the estimated joint distribution and the real joint distribution can be penalized by 
\begin{equation}
\begin{split}
\mathop{\min_{G_{ls}}}\ L_{g}=E_{I\sim p(I)}\sum_{i,j}[1-log\ \sigma(T(G_{ls}(I),I)^{(i,j)})]
\end{split}
\end{equation}
Beyond the optimization of $G_{ls}$ that encouraging model to estimate the right class label at each voxel location independently, this part is taken as the regularization term to regularize the cascade segmentation network to drive it to approximate the real joint distribution of unbalanced atrial targets.

\subsection{Objective Function for Model Learning}
The objective function of JAS-GAN designs for effectively generating reliable results on both the segmentation process and the adversarial training process. Beyond the adversarial training, the adaptive attention cascade network is optimized independently by minimizing the following objective:
\begin{equation}
\begin{split}
\min\limits_{G_{ls}}\ L_{s}=&E_{I\sim p(I)}\ell(Y_{ls},G_{ls}(I))\\
=&\lambda_{1}E_{I\sim p(I)}\ell_{c}(y_{l},\widehat{y}_{l})\\
+&\lambda_{2}E_{I\sim p(I)}\ell_{d}(y_{s},\widehat{y}_{s})
\end{split}
\end{equation}
where $\ell_{c}$ denotes the voxel-wise cross-entropy function while $\ell_{d}$ denotes the Dice-like loss function \cite{Milletari2016V} for addressing the segmentation of small atrial scars. $\lambda_{1}$ and $\lambda_{2}$ are weight parameters used to balance the segmentation losses of the LA and atrial scars.

Then, an adversarial learning is applied to the $G_{ls}$ and $T$ to further regularize the segmentations of LA and atrial scars. We integrate the adversarial learning into the voxel-wise estimations of LA and atrial scars. In this case, the $G_{ls}$ can be learned by minimizing the following objective:
\begin{equation}
\begin{split}
\min\limits_{G_{ls}}\ L_{s}+\lambda_{3}L_{g}
\end{split}
\end{equation}
where $\lambda_{3}$ is the weight parameter used to balance the segmentation loss and the adversarial loss. The $T$ can be learned by directly maximizing $L_{d}$.

\subsection{Network Configuration}
The framework of our proposed JAS-GAN mainly consists of a cascade segmentation network $G_{ls}$ and a joint discriminative network $T$ (Please see the detailed structure of our proposed network architecture in the Appendix). The cascade segmentation network contains three modules of $G_{l}$, $G_{a}$ and $G_{s}$. $G_{l}$ is based on the 2D U-Net \cite{Ronneberger2015U} but uses bilinear interpolation for upsampling. $G_{a}$ is based on a convolutional LSTM network with 32 kernels. $G_{s}$ is based on the residual structure \cite{he2016deep}. It comprises three residual blocks. Each residual block contains three convolution blocks that each convolution block consists of a $5\times 5$ convolution layer with $32$ filters, a batch normalization layer and a ReLU layer. Finally, a $1\times 1$ convolutional layer with sigmoid function is used to predict atrial scars.

The structure of the joint discriminative network $T$ is similar to \cite{hung2019adversarial}. It comprises of 4 convolution layers with $5\times 5$ kernel and \{32, 64, 128, 256\} channels in the stride of two. Each convolution layer is followed by a batch normalization layer and a ReLU layer. Then, a sub-pixel convolution with kernels of 3072 to rescale the output of the last convolution layer to the size of $I$. Finally, a $1\times 1$ convolutional layer with a sigmoid function is used to predict the confidence maps.

\section{Experiments and Results}
\subsection{Data Description}
CMR data were acquired in patients with longstanding persistent AF on a Siemens Magnetom Avanto 1.5T scanner (Siemens Medical Systems, Erlangen, Germany). Transverse navigator-gated 3D LGE CMR \cite{peters2009recurrence} was performed using an inversion prepared segmented gradient echo sequence (TE/TR 2.2ms/5.2ms) 15 minutes after gadolinium administration (Gadovistgadobutrol, 0.1mmol/kg body weight, Bayer- Schering, Berlin, Germany) \cite{haissaguerre1998spontaneous}. The inversion time (TI) was set to null the signal from normal myocardium and varied on a beat-by-beat basis, dependent on the cardiac cycle length \cite{keegan2015dynamic}. Detailed scanning parameters are: 30-34 slices at $(1.4-1.5)\times (1.4-1.5)\times 4\mathrm{mm}^3$, reconstructed to 60-68 slices at $(0.7-0.75)\times(0.7-0.75)\times 2\mathrm{mm}^3$, field-of-view $380\times380\mathrm{mm}^2$. For each patient, prior to contrast agent administration, coronal navigatorgated 3D Roadmap (TE/TR 1ms/2.3ms) data were acquired with the following parameters: 72-80 slices at $(1.6-1.8)\times(1.6-1.8)\times 3.2\mathrm{mm}^3$ , reconstructed to 144-160 slices at $(0.8-0.9)\times(0.8-0.9)\times1.6\mathrm{mm}^3$, field-of-view $380\times 380\mathrm{mm}^2$. LGE CMR was acquired during free-breathing using a crossed-pairs navigator positioned over the dome of the right hemi-diaphragm with navigator acceptance window size of $5\mathrm{mm}$ and CLAWS respiratory motion control \cite{keegan2014improved,keegan2014navigator}. LGE CMR data were collected from 2011-2018 as a retrospective study. In total, 192 scans from 115 subjects including 97 pre-ablation and 95 post-ablation scans were used in this study (All subjects gave their informed consent for inclusion before they participated in the study with approval from the local institutional review board in accordance with the Declaration of Helsinki (Ethics approval reference number: 10/H0701/112, CMR Unit, Royal Brompton Hospital)). Manual segmentations of the LA and proximal PVs and atrial scars had been done by a physician with $>$ 3 years of experience and specialized in LGE CMR. A second senior radiologist ($>$25 years of experience and specialized in cardiac MRI) confirmed the manual segmentations. The results confirmed by the radiologist were chosen as the ground truth for experiments. The LA label is the LA epicardium (LA wall and LA cavity). Because the atrial scars are located in LA wall, the atrial scars label is encapsulated in the LA label.

\begin{table}[!hbtp]
 \setlength{\abovecaptionskip}{0pt} 
 \setlength{\belowcaptionskip}{0pt} 
 \caption{\scriptsize{Quantitative results of unbalanced atrial targets segmentation on pre-ablation, post-ablation and pre-\&post-ablations in terms of DSC, JI, ASD and NMI. Results are presented in the form of mean $\pm$ standard deviation. Abbreviations: DSC, Dice Similarity Coefficient;  JI, Jaccard Index; ASD, Average Surface Distance; NMI, Normalized Mutual Information.}}
 \centering
 \scalebox{.55}{
 \begin{tabular}{cccccc}
 \addlinespace
 \toprule
 Target &  Pre-/Post-ablation & \multicolumn{1}{c}{DSC} & \multicolumn{1}{c}{JI} & \multicolumn{1}{c}{ASD (mm)} &\multicolumn{1}{c}{NMI}\\ \midrule
 \multirow{3}*{LA and PVs} & Pre-ablation & $0.948\pm0.012$ &$0.901 \pm 0.022$    &$0.869 \pm 0.381$ & $0.847\pm0.026$ \cr 
                           &Post-ablation & $0.944\pm0.017$ &$0.893 \pm 0.029$    &$0.966 \pm 0.522$ & $0.841\pm0.034$ \cr
                           &Pre-\&Post-ablations &$0.946\pm0.015$  &$0.897 \pm 0.026$  & $0.918\pm 0.460$ & $0.844\pm0.030$\\ \midrule
 \multirow{3}*{Atrial scars} &Pre-ablation & $0.809\pm 0.064$ &$0.684 \pm 0.087$  &$0.483 \pm 0.280$ & $0.703\pm0.077$ \cr 
                             &Post-ablation & $0.832\pm0.051$ &$0.716 \pm 0.073$  &$0.394 \pm 0.158$ & $0.722\pm0.067$ \cr
                             &Pre-\&Post-ablations  &$0.821\pm 0.059$ &$0.700 \pm 0.082$   & $0.439\pm 0.232$ & $0.713\pm0.073$ \\ \midrule
 \end{tabular}
 }
 \label{table:segmentation performance}
\end{table}

\subsection{Implementation Details}
We performed data normalization on the whole 3D volume for all experiments. In addition, because of the small proportion of positive pixels per axial slice, it is very ineffective to train the segmentation model on the entire LGE-MRI data directly. To relieve this, smaller patches of $256\times 256$ which contained positive and negative pixels centered on the raw LGE CMR image were generated as inputs. 

We randomly divided our dataset into a training set (116 scans from 77 patients) and a testing set (76 scans from 38 patients with 38 pre-ablation and 38 post-ablation scans) for all experiments. The divided strategy for the dataset was that all scans from each unique patient were only in one of the training or testing sets.

We used the Adam method to perform the optimization for the cascade segmentation network with a decayed learning rate (the initial learning rate was set to 0.001 with a decay rate of $0.99$). The optimizer used for the joint discriminative network was Adam with a fixed learning rate of 0.0001. We used the current statistics of batch normalization for the both training and testing. In addition, to stabilize the training of GAN, we used the feature matching \cite{Salimans2016Improved} for adversarial loss. The coefficients of $\lambda_{1}$ and $\lambda_{2}$ used to balance the two segmentation losses, were automatically learned based on the strategy of uncertainty \cite{kendall2018multi}. The coefficient of $\lambda_{3}$ used to balance the segmentation loss and adversarial loss, was set to a fixed value of 0.1.

Our deep learning model was implemented using Tensorflow $1.2.1$ on an Ubuntu $16.04$ machine and was trained and tested using an Nvidia RTX 8000 GPU (48GB GPU memory).

\begin{figure}[!hbtp]
\begin{center}
   \includegraphics[width=1\linewidth]{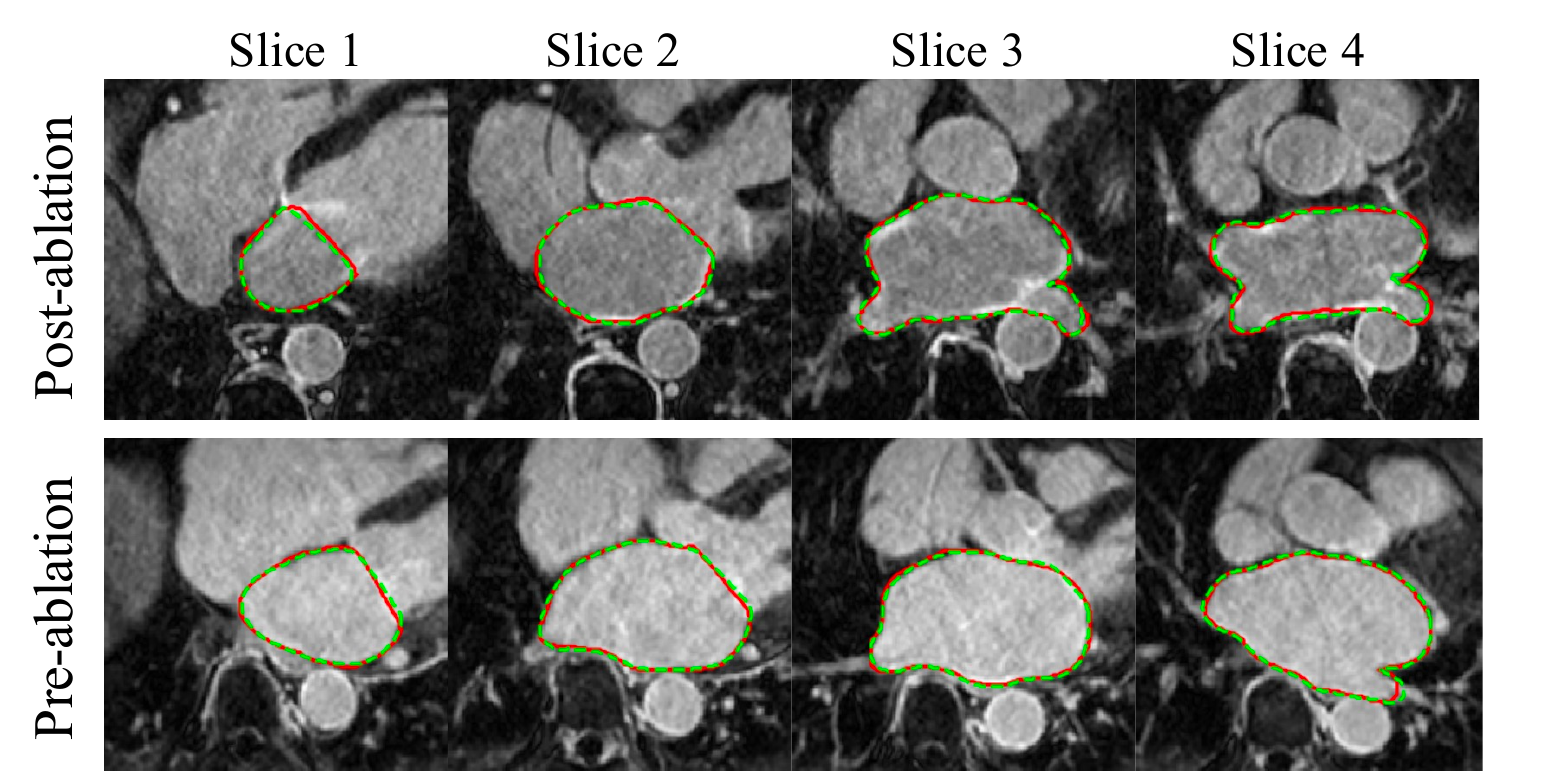}
\end{center}
\caption{Qualitative visualization of the segmentation for the LA in pre-ablation and post-ablation cases. Each estimated segmentation is represented as a dashed green counter while the red contour denotes its corresponding manually delineated ground truth.}
\label{fig:atrium_vs} 
\end{figure}

\begin{figure}[!hbtp]
\begin{center}
   \includegraphics[width=1\linewidth]{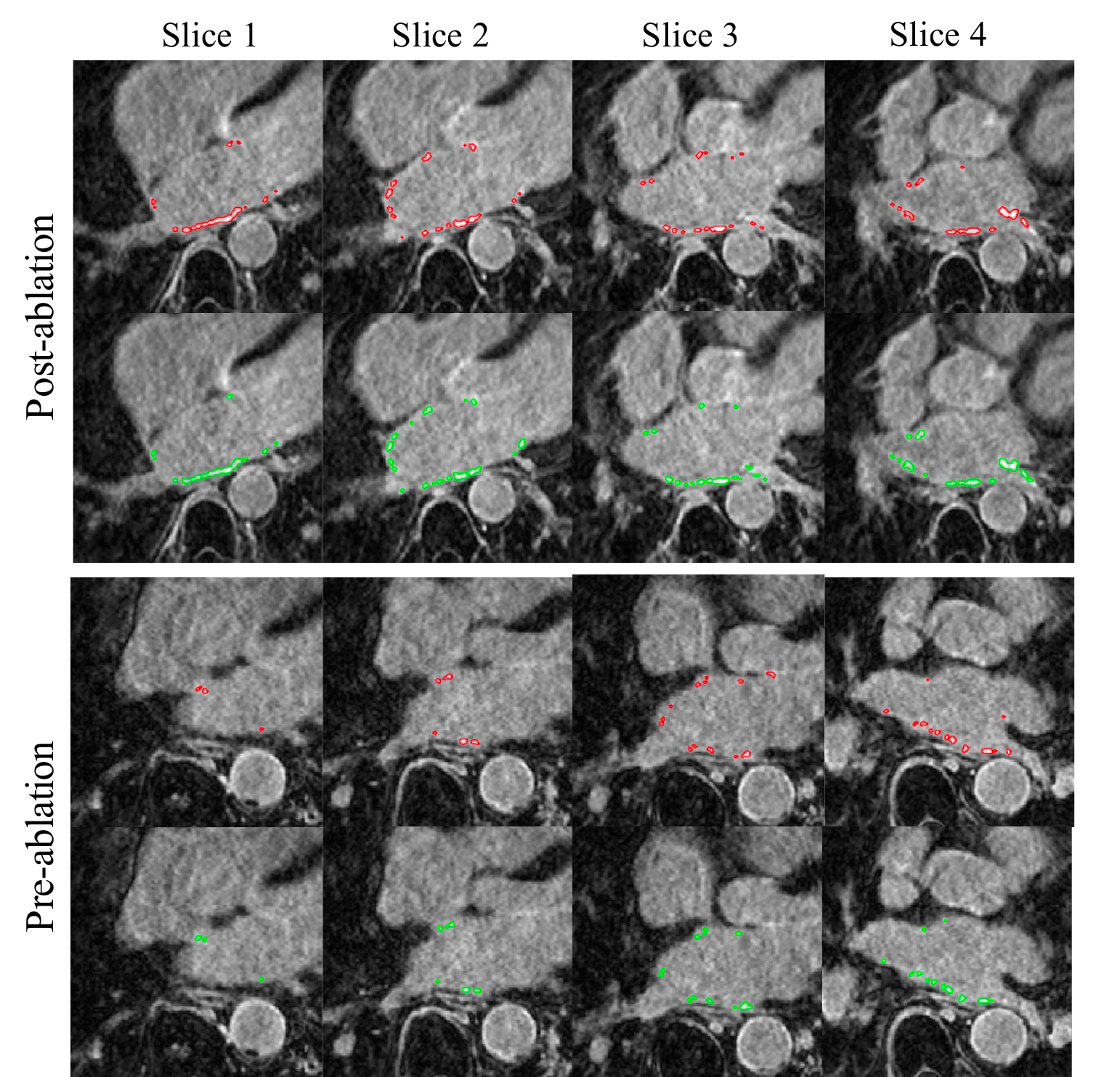} 
\end{center}
\caption{Qualitative visualization of the segmentation for the atrial scars in pre-ablation and post-ablation cases. Each estimated segmentation is represented as a dashed green counter while the red contour denotes its corresponding manually delineated ground truth.}
\label{fig:scar_vs} 
\end{figure}

\begin{figure}[!hbtp]
\begin{center}
\scalebox{.9}{
   \includegraphics[width=1\linewidth]{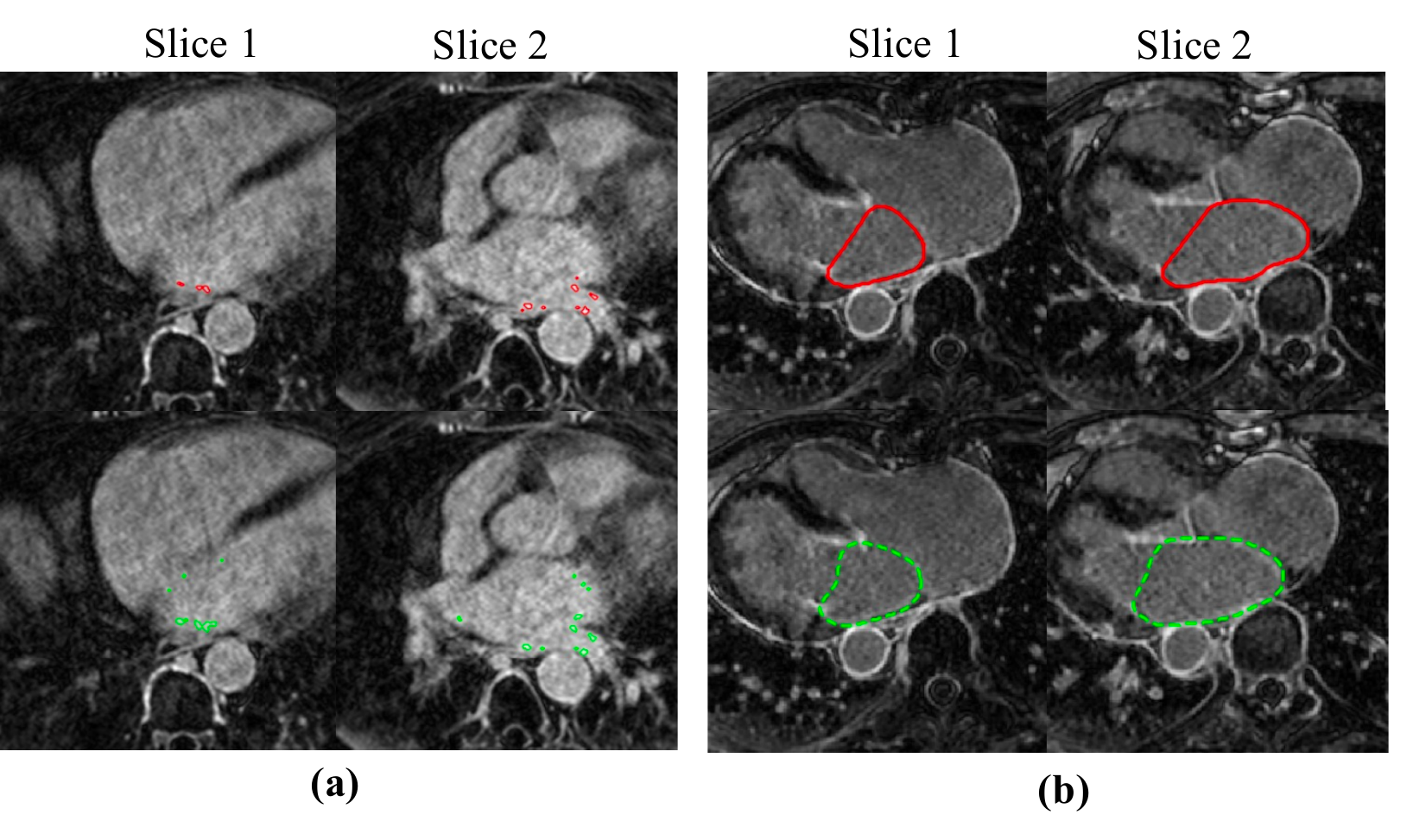}
   }
\end{center}
\caption{Qualitative visualization of the segmentation for LA and atrial scars in worst cases. Each estimated segmentation is represented as a dashed green counter (bottom raw) while the red contour denotes its corresponding manually delineated ground truth (top raw). (a) Qualitative visualization of the segmentation for atrial scars. (b) Qualitative visualization of the segmentation for LA.}
\label{fig:poor} 
\end{figure}

\subsection{Evaluation Criteria}
To evaluate the segmentation performance of our proposed JAS-GAN, we used region-based metrics \cite{dice1945measures,taha2015metrics}, e.g., the Dice Similarity Coefficient (DSC) and the Jaccard Index (JI), to validate the predicted segmentation map against the manually defined ground-truth. We also used a surface-based metric called Average Surface Distance (ASD) to provide the distance in $\mathrm{mm}$ to quantify the accuracy of the predicted mesh ($S$) compared to the ground-truth mesh ($S^\prime$)  \cite{taha2015metrics}. We further adopted the Normalized Mutual Information (NMI) to measure the similarity between the estimated segmentation maps and the ground truth \cite{taha2015metrics}. In addition, the segmentation performance of our proposed JAS-GAN was further evaluated by the over-segmentation rate (OSR) and the under-segmentation rate (USR), which are defined as $USR = FN/(TP+FN)$ and $OSR = FP/(TP+FN)$ \cite{miao2018image}, where TP, FP and FN denote the True Positive, the False Positive, and the False Negative, respectively. They are defined as the number of voxels correctly identified as positive for target, the number of voxels incorrectly identified as positive for target, and the number of voxels incorrectly identified as negative for target, respectively. TP, FP and FN were calculated while considering all voxels in a 3D volume.

\begin{table}[!hbtp]
 \setlength{\abovecaptionskip}{0pt} 
 \setlength{\belowcaptionskip}{0pt} 
 \caption{\scriptsize{Ablation results comparison for JAS-GAN in terms of DSC, JI, ASD and NMI. The results are presented in the form of mean $\pm$ standard deviation. Abbreviations: EDN, baseline based on encoder-decoder network for LA segmentation; RN, baseline based on residual network for atrial scars segmentation; AC, adaptive attention cascade; T: joint discriminative network; DSC, Dice Similarity Coefficient;  JI, Jaccard Index; ASD, Average Surface Distance; NMI, Normalized Mutual Information.}}
 \centering
 \scalebox{.5}{
\begin{tabular}{cccccc}
\addlinespace
\toprule
Target &  Methods & \multicolumn{1}{c}{DSC} & \multicolumn{1}{c}{JI} & \multicolumn{1}{c}{ASD (mm)} &\multicolumn{1}{c}{NMI}\\ \midrule
 \multirow{4}*{LA and PVs} &EDN & $0.930\pm0.023$ &$0.870 \pm 0.039$ &$1.20 \pm 0.567$ & $0.812\pm0.046$ \cr 
                           &EDN + AC & $0.936\pm0.021$ &$0.881 \pm 0.036$ &$1.09 \pm 0.703$ & $0.823\pm0.040$ \cr 
                           &EDN + AC + T (JAS-GAN) &$0.946\pm0.015$  &$0.897 \pm 0.026$  & $0.918\pm 0.460$ & $0.844\pm0.030$ \\ \midrule      
 \multirow{4}*{Atrial scars} &RN & $0.778\pm0.074$ &$0.643 \pm 0.092$  &$0.654 \pm 0.347$ & $0.660\pm0.083$ \cr 
                             &RN + LA & $0.784\pm0.064$ &$0.649 \pm 0.084$  &$0.567 \pm 0.270$ & $ 0.667\pm0.075$ \cr 
                             &RN + AC & $0.810\pm0.061$ &$0.686 \pm 0.082$  &$0.489 \pm 0.251$ & $0.701\pm0.072$ \cr            
                             &RN + AC + T (JAS-GAN) &$0.821\pm 0.059$ &$0.700 \pm 0.082$   & $0.439\pm 0.232$ & $0.713\pm0.073$ \\ \midrule
\end{tabular}
 }
 \label{table:ablation results} 
\end{table}

\subsection{Segmentation Performance of JAS-GAN.}
Quantitative analysis: Table \ref{table:segmentation performance} summarizes the quantitative segmentation results of JAS-GAN grouped by the pre-ablation, post-ablation and pre-\&post-ablations. Despite the challenges in segmenting the LA and atrial scars from LGE CMR scans, our proposed JAS-GAN still achieved high segmentation accuracy in terms of DSC, JI, ASD and NMI for both LA and atrial scars in pre-ablation, post-ablation and pre-\&post-ablations. We had further performed the statistical tests (t-test) to show the statistical differences between the pre-ablation and the post-ablation. The calculated lowest P-values of $0.248$ and $0.093$ in terms of DSC, JI, ASD and NMI for LA and atrial scars demonstrated there were no significant differences between pre-ablation and post-ablation. Furthermore, we had investigated the inter-observation variability and the inter-observer agreement from two manual segmentations of LA and atrial scars. We provided 12 cases selected from the testing data (6 pre-ablation scans and 6 post-ablation scans) for two experts to manually label the LA and atrial scars independently. We had followed the \cite{joskowicz2019inter} to use the mean (1-DSC) and mean DSC with ranges to measure the inter-observer variability and the inter-observer agreement based on the mean volume overlap variability values and the mean volume overlap values, respectively. The inter-observer variabilities for the LA and atrial scars  are  0.082 [-0.009,0.007] and 0.291 [-0.072, 0.063], respectively. The inter-observer agreements for the LA and atrial scars were 0.918 [-0.007,0.009] and 0.709 [-0.063, 0.072], respectively.  Compared with the inter-observer agreements, JAS-GAN has achieved higher segmentation accuracies for the unbalanced atrial targets. These results indicated the ability of JAS-GAN in handling the automated and accurate segmentations of LA and atrial scars. 

Qualitative analysis: Fig. \ref{fig:atrium_vs} and Fig. \ref{fig:scar_vs} show the qualitative segmentation results of JAS-GAN compared to the ground truth for the selected slices of pre-ablation and post-ablation. One can see that our proposed JAS-GAN has the ability to handle the shape and size variations of the LA and small atrial scars. We also provided the qualitative segmentation results of LA and atrial scars for the worst cases. The studied cohort is a difficult AF cohort that the patients have severe arrhythmia during the MRI scanning, the blurry LA boundary and indistinguishable atrial scars are major reasons for the worst segmentation results as shown in Fig. \ref{fig:poor} (a) and (b). However, one also can see that our proposed JAS-GAN still achieves general segmentation results in visual. We will collect more LGE CMR data for model learning to overcome the blurry LA boundary and indistinguishable atrial scar.

\begin{figure}[!hbtp]
\begin{center}
\scalebox{.85}{
   \includegraphics[width=1\linewidth]{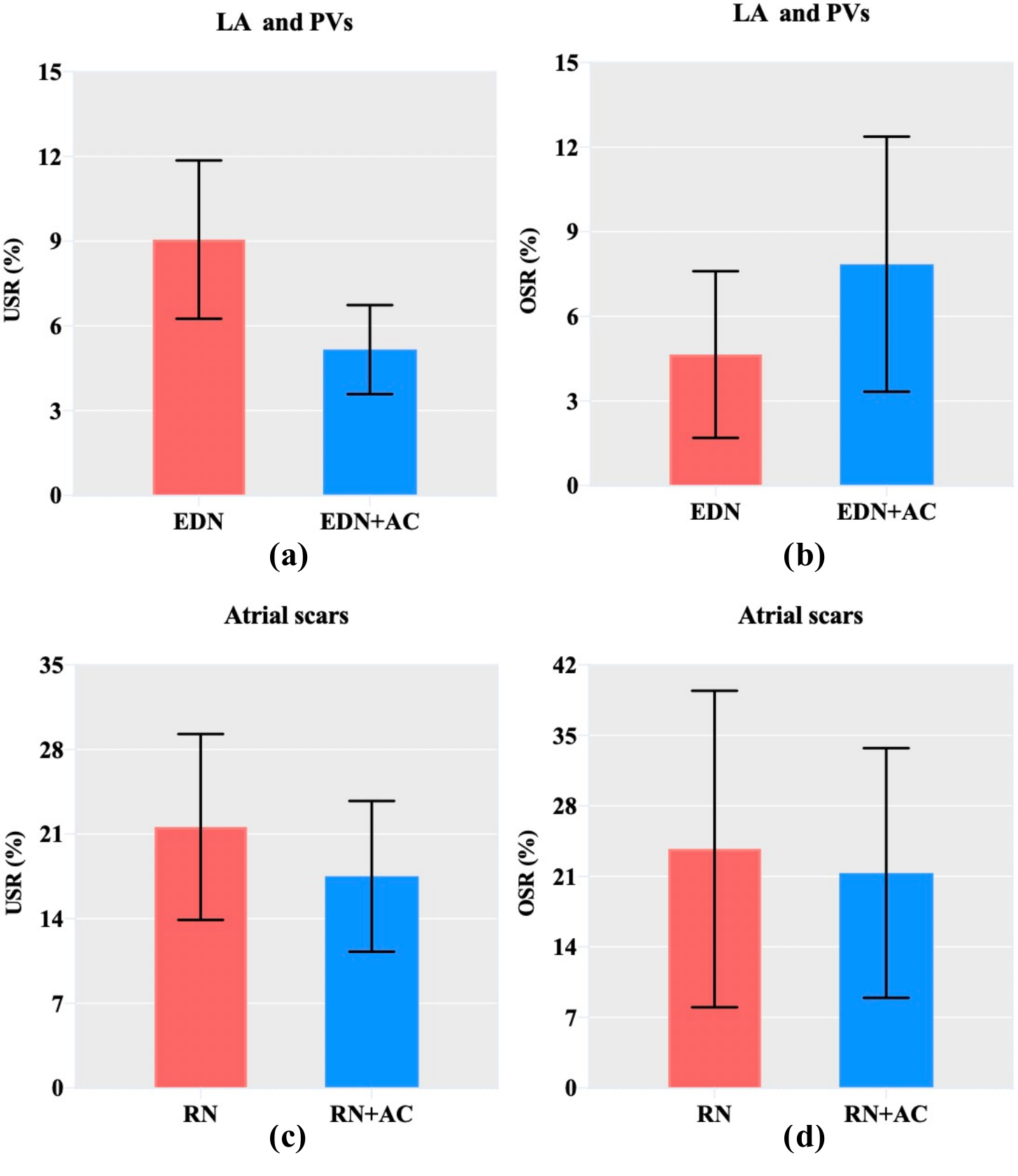}
   }
\end{center}
\caption{Analysis of over-segmentation rate (OSR) and under-segmentation rate (USR) for LA and atrial scars (Black bars denote standard deviation). (a) and (b) denote under-segmentation rate and over-segmentation rate comparison for the LA segmentation with/without using cascade connection. (c) and (d) denote under-segmentation rate and over-segmentation rate comparison for the atrial scars segmentation with/without using cascade connection. Abbreviations: EDN, baseline based on a encoder-decoder for LA segmentation; RN, baseline based on a residual network for atrial scars segmentation; AC, adaptive attention cascade.}
\label{fig:OSR_USR} 
\end{figure}

\subsection{Ablation Analysis of JAS-GAN.}
The effectiveness of the adaptive attention cascade network and the joint discriminative network was extensively analysed with ablation experiments. Firstly, the baselines based on the encoder-decoder network $G_{l}$ (EDN) and the residual network $G_{s}$ (RN) were performed for the LA and atrial scars segmentations, respectively. Then, we used the LA segmentation results of EDN to further define a region of interest (ROI) in the input image for further RN-based scar segmentation (RN+LA). Next, we constructed a cascade network that EDN and RN were cascaded by the adaptive attention cascade (AC) to perform the joint segmentations of LA and atrial scars in an end-to-end manner (EDN + AC for LA, RN + AC for atrial scars). Finally, based on the cascade network, we added the joint discriminative network T for adversarial regularization (EDN + AC + T for LA, RN + AC + T for atrial scars).

\textit{1) Effectiveness of adaptive attention cascade network}: Adaptive attention cascade network leverages an adaptive attention cascade to automatically correlate the segmentation tasks of LA and atrial scars for their joint segmentations. The adaptive attention cascade makes the segmentation model try to produce the over-segmented LA rather than to produce the under-segmented LA for focusing on the small atrial scars roughly. As the results are shown in Table \ref{table:ablation results}, adaptive attention cascade can both improve the segmentation performance of LA and atrial scars in terms of $DSC$, $JI$, $ASD$ and $NMI$ (EDN + AC vs. EDN, RN + AC vs. RN). One also can see that the improvement of segmentation accuracy for atrial scars was limited while using the LA segmentation output to define ROI in the image for further scar segmentation (RN + LA vs. RN). The reason is that the under-segmented LA can weaken the atrial scars partially or completely while using two-stage segmentation. The improvements of adaptive attention cascade had been demonstrated to be statistically significant based on t-tests (P-values $<$ 0.05). Fig. \ref{fig:OSR_USR} summarizes the over-segmented and under-segmented results for estimated LA and atrial scars. The Fig. \ref{fig:OSR_USR} (a) and (b) show that EDN with AC achieved lower $USR$ and higher $OSR$ compared to EDN for the estimated LA, which illustrated that EDN with AC tries to produce over-segmented LA to pay attention to the small atrial scars. Fig. \ref{fig:OSR_USR} (c) and (d) show that RN with AC achieved the lower $USR$ and $OSR$ compared to RN for the segmentation of small atrial scars, which indicated that adaptive attention cascade leverages estimated LA to constrain the learnable area of small atrial scars for their accurate identification.

\begin{figure}[!hbtp]
\begin{center}
   \includegraphics[width=1\linewidth]{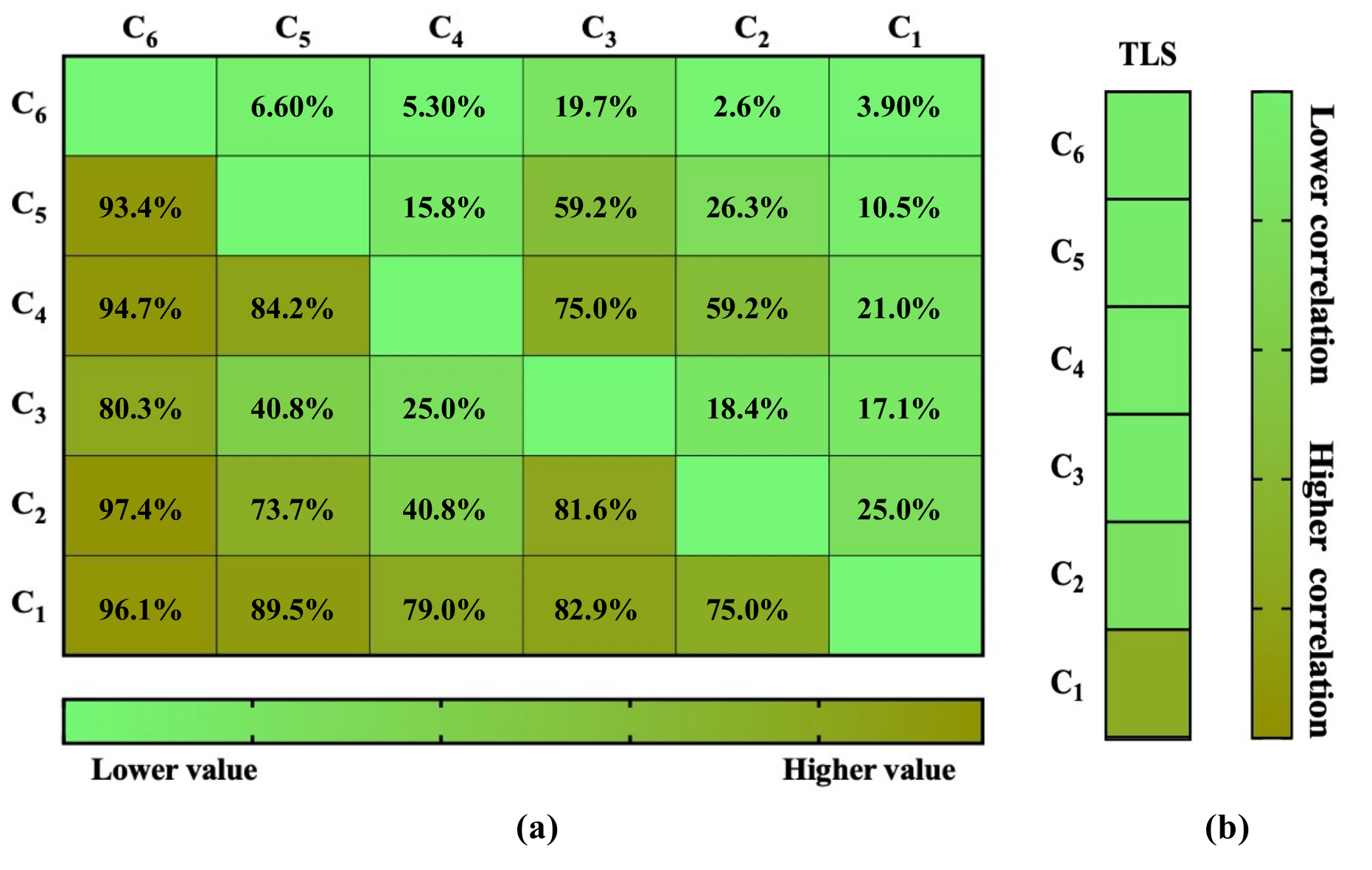}
\end{center}
\caption{Analysis of different cascade information for correlating the segmentation tasks of unbalanced atrial targets. (a) Pairwise tournament matrix for measuring the superiority of one cascade information relative to others for the correlation extent between the segmentation tasks of LA and atrial scars. $C_{1}$ obtains the best superiority for the correlation extent between the two segmentation tasks. (b) First-order task affinity matrix for measuring the correlation between the two segmentation tasks for different cascade information. $C_{1}$ achieves the best correlation between the two segmentation tasks. Abbreviations: $C_{1}$, LA probability map; $C_{6}$, $C_{5}$, $C_{4}$, $C_{3}$ and $C_{2}$, information output by the encoder, the first up-sampling block of decoder, the second up-sampling block of decoder, the third up-sampling block of decoder and the fourth up-sampling block of decoder in LA segmentation network, respectively; TLS: segmentation tasks of LA and atrial scars }
\label{fig:cascade_correlation} 
\end{figure}

\textit{2) Effectiveness of the joint discriminative network}: As the experiment results are shown in Table \ref{table:ablation results}, compared with EDN + AC and RN + AC, JAS-GAN achieved better segmentation results for LA and atrial scars across all evaluation metrics, which indicated that the adversarial regularization achieved by the joint discriminative network is effective to improve the segmentation performance of adaptive attention cascade network. In addition, the improvements of the joint discriminative network had been demonstrated to be statistically significant (P-values $<$ 0.05) based on t-tests. 

\begin{table}[!hbtp]
 \setlength{\abovecaptionskip}{0pt} 
 \setlength{\belowcaptionskip}{0pt} 
 \caption{\scriptsize{Performance comparison of different cascade operations for atrial scars segmentation in terms of DSC, JI, ASD AND NMI. The results are presented in the form of mean $\pm$ standard deviation. Abbreviations: DSC, Dice Similarity Coefficient;  JI, Jaccard Index; ASD, Average Surface Distance; NMI, Normalized Mutual Information; $O_{a}$, element-wise add operation; $O_{p}$, element-wise product operation; $O_{c}$, concatenation operation; $O_{ac}$, our used adaptive attention operation.}}
 \centering
 \scalebox{.7}{
    \begin{tabular}{ccccc}
    \toprule
    Methods & \multicolumn{1}{c}{DSC} & \multicolumn{1}{c}{JI} & \multicolumn{1}{c}{ASD (mm)} &\multicolumn{1}{c}{NMI}\\ \midrule
     $O_{a}$ & $0.799\pm0.071$ &$0.670 \pm 0.093$  &$0.456 \pm 0.232$ & $0.687\pm0.080$ \cr 
     $O_{p}$ & $0.810\pm 0.064$ &$0.685 \pm 0.086$  &$0.447 \pm 0.249$ & $0.700\pm0.075$ \cr
     $O_{c}$ & $0.813\pm 0.061$ &$0.689 \pm 0.083$  &$0.461 \pm 0.258$ & $0.703\pm0.072$ \cr
     $O_{ac}$ &$0.821\pm 0.059$ &$0.700 \pm 0.082$   & $0.439\pm 0.232$ & $0.713\pm0.073$ \\ \midrule
    \end{tabular}
 }
 \label{table:correlation results} 
\end{table}

\subsection{Analysis of Adaptive Attention Cascade Connection.}
To analyse the feasibility of adaptive attention cascade connection, we performed extra experiments to validate the effectiveness of our used cascade information (LA probability map) and cascade operation (adaptive attention). In our proposed cascade framework, we used the estimated LA probability map, which represents complete decoding for the LA feature information, as the cascade information to correlate the segmentation tasks of LA and atrial scars. To demonstrate the superior of the LA probability map, we further investigated the influence of LA information with different decoding levels for correlating the segmentation tasks of LA and atrial scars. In our experiment, in addition to the LA probability map ($C_{1}$), we also used extra five kinds of LA information with different decoding levels from the LA segmentation network (encoder-decoder) as feedforward information to correlate the segmentation tasks of LA and atrial scars. They were the information output by the encoder ($C_{6}$), the first up-sampling block of the decoder ($C_{5}$), the second up-sampling block of the decoder ($C_{4}$), the third up-sampling block of the decoder ($C_{3}$) and the fourth up-sampling block of the decoder ($C_{2}$). We followed the \cite{zamir2018taskonomy} to construct a pairwise tournament matrix to measure the superiority of each information to correlate the segmentation tasks of LA and atrial scars compared to other information. As the constructed pairwise tournament matrix are shown in Fig. \ref{fig:cascade_correlation} (a), the element at $(i,j)$ of pairwise tournament matrix is the percentage of data in a test set $D_{test}$, on which $C_{i}$ correlates the segmentation tasks of LA and atrial scars ($TLS$) better than $C_{j}$ did (i.e. $D_{C_{i}\rightarrow TLS}(I) \> D_{C_{j}\rightarrow TLS}(I)$). Based on the pairwise tournament matrix, we further obtained the affinity matrix as shown in Fig. \ref{fig:cascade_correlation} (b), where each value represents the correlation between the two segmentation tasks achieved by the corresponding cascade information. As the results are shown in Fig. \ref{fig:cascade_correlation} (b), the cascade information of the estimated LA probability map obtained the best correlation for the two segmentation tasks. 

To demonstrate that our used cascade operation of adaptive attention ($O_{ac}$) is  effective to segment small atrial scars, we further compared it to the pixel-wise add operation ($O_{a}$) which tries to use the LA segmentation output to enhance the LA region of image for subsequent segmentation of atrial scars, general attention operation with pixel-wise product ($O_{p}$) which tries to use the LA segmentation output to define ROI in input image for atrial scars segmentation with end-to-end model optimization, and direct concatenation operation ($O_{c}$) of the estimated LA and the input image. As the summarized results are shown in Table \ref{table:correlation results}, our used adaptive attention cascade achieved better segmentation results. Furthermore, the improvements had been demonstrated to be statistically significant based on t-tests (P-values $<$ 0.05). Those indicated the  effectiveness of adaptive attention for the segmentation of small atrial scars. 

\begin{figure}[!hbtp]
\begin{center}
\scalebox{.75}{
   \includegraphics[width=1\linewidth]{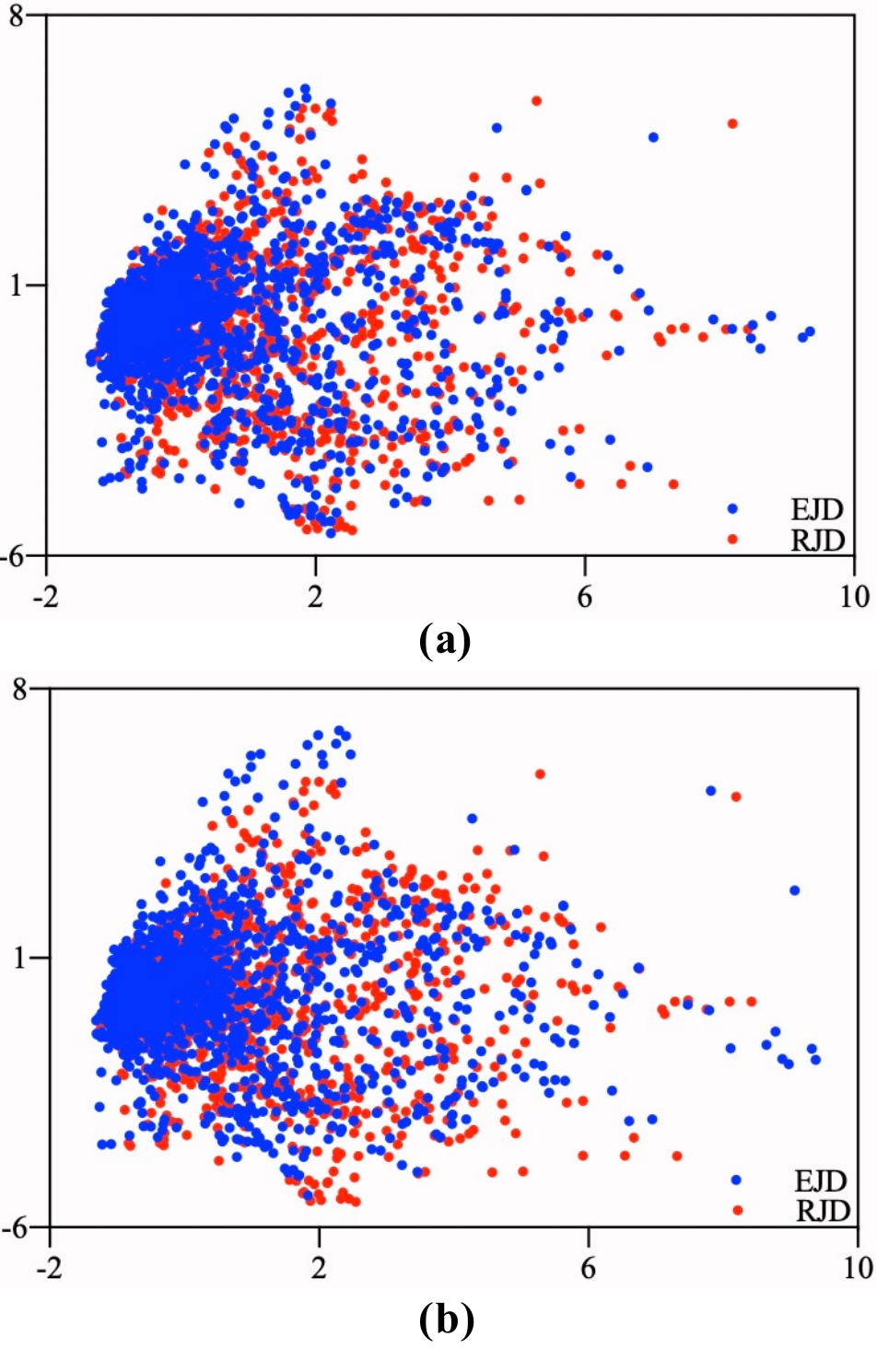}
   }
\end{center}
\caption{Principal components analysis (PCA) based visualization of joint distribution for LA and atrial scars. The x-axis represents the direction with the largest variance of the data before PCA operation while the y-axis represents the direction orthogonal to the x-axis with the largest variance of the data before PCA operation. (a) The joint distribution estimated by JAS-GAN with joint discriminative network and real ones.  (b) The joint distribution estimated by JAS-GAN without joint discriminative network and real ones. Abbreviations: EJD, estimated joint distribution; RJD: real joint distribution).}
\label{fig:joint_distribution} 
\end{figure}

\subsection{Analysis of Joint Discriminative Network.}
In our proposed JAS-GAN, the joint discriminative network is used to further transform the semantic segmentation of pixel-level classification for the unbalanced targets into the joint pixel-level identification of unbalanced targets. It mainly utilizes the adversarial regularization to force the estimated joint distribution of LA and atrial scars produced by the cascade segmentation network to match the real ones. To demonstrate that the joint discriminative network has the ability to improve the consistency of joint distribution for LA and atrial scars, we had made a visualization of the estimated joint distribution (EJD) and the real ones (RJD) based on the principal components analysis (PCA) for visually assessing the matching degree of EJD and RJD as shown in Fig. \ref{fig:joint_distribution}, where the data points in the estimated joint distribution and the data points in the real ones were in a one-to-one correspondence. We also provided a quantitative distance between the EJD and the RJD based on the mean Euclidean distance of the corresponding points in the 2-dimensional coordinate system (Fig. \ref{fig:joint_distribution}).  The calculated distance between the EJD by JAS-GAN with joint discriminative network and the RJD was $0.211$ while the calculated distance between the EJD by JAS-GAN without joint discriminative network and the RJD was 0.304. The qualitative visualization and the quantitative distance both denoted the JAS-GAN with joint discriminative network achieved a more consistent joint distribution between the estimated results and the real ones compared to JAS-GAN without joint discriminative network.

\begin{table}[!hbtp]
 \setlength{\abovecaptionskip}{0pt} 
 \setlength{\belowcaptionskip}{0pt} 
 \caption{\scriptsize{LA segmentation performance comparison between the different architectures (Segnet, 3D Densenet, 2D U-Net, 3D U-Net, MTL, MVTT and JAS-GAN) and the methods aiming to tackle the imbalance issue (Tversky loss and surface loss) in terms of DSC, JI, ASD AND NMI. The results are presented in the form of mean $\pm$ standard deviation. Abbreviations: DSC, Dice Similarity Coefficient;  JI, Jaccard Index; ASD, Average Surface Distance; NMI, Normalized Mutual Information.}}
 \centering
 \scalebox{.65}{
    \begin{tabular}{cccccc}
    \addlinespace
    \toprule
    Methods & \multicolumn{1}{c}{DSC} & \multicolumn{1}{c}{JI} & \multicolumn{1}{c}{ASD (mm)} &\multicolumn{1}{c}{NMI}\\ \midrule
     Segnet      & $0.915\pm0.023$   &$0.844 \pm 0.039$  &$1.61 \pm 0.615$ & $0.779\pm0.044$ \cr 
     3D Densenet & $0.921\pm 0.017$  &$0.855 \pm 0.029$  &$1.48 \pm 0.554$ & $0.790\pm0.032$ \cr 
     2D U-Net     & $0.928\pm 0.021$  &$0.867 \pm 0.036$  &$1.21 \pm 0.553$ & $0.806\pm0.041$ \cr 
     3D U-Net     & $0.926\pm 0.018$  &$0.863 \pm 0.032$  &$1.29\pm 0.715$  & $0.803\pm0.036$ \cr 
     MTL         & $0.934\pm 0.020$  &$0.876 \pm 0.035$  &$1.10 \pm 0.413$ & $0.818\pm0.039$ \cr 
     Tversky Loss & $0.936\pm 0.021$ &$0.881 \pm 0.035$  &$1.12 \pm 0.667$ & $0.825\pm0.040$ \cr
     Surface Loss & $0.938\pm 0.020$ &$0.884 \pm 0.034$  &$1.17 \pm 0.825$ & $0.829\pm0.038$ \cr 
     MVTT        & $0.938\pm 0.023$ &$0.884 \pm 0.039$  &$1.07 \pm 0.612$ & $0.827\pm0.047$ \cr 
    JAS-GAN      & $\mathbf{0.946\pm 0.015}$ &$\mathbf{0.897 \pm 0.026}$  &$\mathbf{0.918 \pm 0.460}$ & $\mathbf{0.844\pm0.030}$ \\ \midrule
    \end{tabular}
 }
 \label{table:LA_comparison} 
\end{table}

\begin{table}[!hbtp]
 \setlength{\abovecaptionskip}{0pt} 
 \setlength{\belowcaptionskip}{0pt} 
 \caption{\scriptsize{Scar segmentation performance comparison between the different architectures (Segnet, 3D Densenet, 2D U-Net, 3D U-Net, MVTT and JAS-GAN), the two-phase methods (2SD and Ostu) and the methods aiming to tackle the imbalance issue (Tversky loss and surface loss) in terms of DSC, JI, ASD AND NMI. The results are presented in the form of mean $\pm$ standard deviation. Abbreviations: DSC, Dice Similarity Coefficient;  JI, Jaccard Index; ASD, Average Surface Distance; NMI, Normalized Mutual Information.}}
 \centering
 \scalebox{.65}{
    \begin{tabular}{cccccc}
    \addlinespace
    \toprule
    Methods & \multicolumn{1}{c}{DSC} & \multicolumn{1}{c}{JI} & \multicolumn{1}{c}{ASD (mm)} &\multicolumn{1}{c}{NMI}\\ \midrule
     2SD         & $0.405\pm0.210$   &$0.278 \pm 0.179$  &$2.37 \pm 1.75$ & $0.317\pm0.183$ \cr 
     Ostu        & $0.442\pm 0.232$  &$0.313 \pm 0.203$  &$7.01 \pm 22.3$ & $0.351\pm0.204$ \cr 
     Segnet      & $0.537\pm0.119$   &$0.376 \pm 0.107$  &$1.63 \pm 0.839$ & $0.403\pm0.106$ \cr 
     3D Densenet & $0.625\pm 0.119$  &$0.464 \pm 0.117$  &$0.979 \pm 0.598$ & $0.486\pm0.111$ \cr 
     2D U-Net     & $0.772\pm 0.078$  &$0.635 \pm 0.098$  &$0.602 \pm 0.298$ & $0.653\pm0.085$ \cr 
     3D U-Net     & $0.754\pm 0.083$  &$0.612 \pm 0.110$  &$0.631\pm 0.406$  & $0.633\pm0.089$ \cr 
     Tversky Loss & $0.780\pm 0.070$ &$0.645 \pm 0.088$  &$0.616 \pm 0.290$ & $0.664\pm0.079$ \cr
     Surface Loss & $0.792\pm 0.070$ &$0.661 \pm 0.090$  &$0.589 \pm 0.329$ & $0.678\pm0.078$ \cr 
     MVTT        & $0.811\pm 0.061$ &$0.687 \pm 0.083$  &$0.502 \pm 0.289$ & $0.701\pm0.074$ \cr 
    JAS-GAN      & $\mathbf{0.821\pm 0.059}$ &$\mathbf{0.700 \pm 0.081}$  &$\mathbf{0.439 \pm 0.232}$ & $\mathbf{0.713\pm0.073}$ \\ \midrule
    \end{tabular}
 }
 \label{table:scar_comparison} 
\end{table}

\subsection{Performance Comparison with Other Methods.}
The performance of JAS-GAN had further been demonstrated by comparing it with the widely used methods and the state-of-the-art methods. For the segmentations of LA and atrial scars, we compared the segmentation performance of JAS-GAN to the 2D U-Net \cite{Ronneberger2015U}, 3D U-Net \cite{cciccek20163d}, 3D DenseNet \cite{bui20173d}, SegNet \cite{badrinarayanan2017segnet}, the method (MVTT) proposed by yang et al.\cite{yang2020simultaneous} and two methods aiming to tackle the imbalance issue (Tversky loss \cite{salehi2017tversky} and surface loss \cite{kervadec2019boundary}). We also compare the LA segmentation performance of JAS-GAN to the method (MTL) proposed by Chen et al.\cite{chen2018multi}. Table \ref{table:LA_comparison} and Table \ref{table:scar_comparison} summarizes the experiment results. 

For the LA segmentation, as the experiment results are shown in Table \ref{table:LA_comparison}, JAS-GAN achieved higher segmentation performance with improved $DSC$, $JI$, $NMI$ and reduced $ASD$ compared to other methods. The reason behind this is that we make full use of adaptive attention cascade connection and adversarial regularization to promote the performance of the encoder-decoder structure.

For the segmentation of atrial scars, as the experiment results are shown in Table \ref{table:scar_comparison}, our proposed JAS-GAN outperformed the compared methods in terms of $DSC$, $JI$, $NMI$ and $ASD$.  The segmentation accuracy obtained by the widely used deep learning methods was limited. This is because they have no suitable mechanism to segment very small atrial scars. The Tversky loss and the Surface loss can effectively deal with imbalance problems. However, only relying on the loss function to deal with the problem of unbalanced target segmentation is still limited to the improvement of segmentation accuracy. The evaluation metrics also illustrated that our proposed JAS-GAN resulted in a more effective architecture for atrial scars segmentation compared to MVTT.

\begin{figure}[!htp]
\begin{center}
\scalebox{.99}{
   \includegraphics[width=1\linewidth]{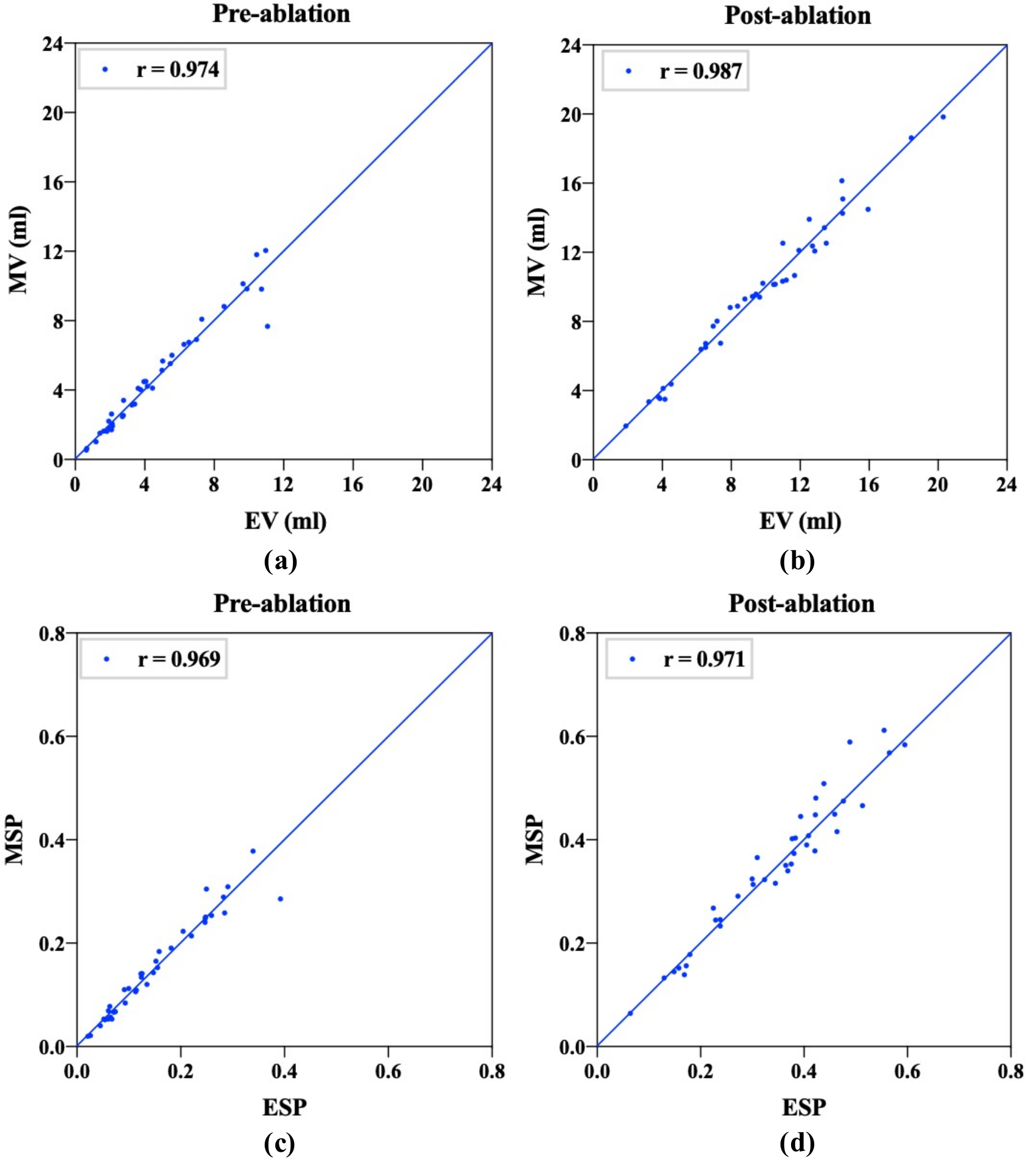}
   }
\end{center}
\caption{The correlation analysis for JAS-GAN. (a) and (b) The high correlations between estimated scar volume and ground truth for pre-ablation and post-ablation, respectively. (c) and (d) The high correlations between estimated scar percentage and manual segmentation for pre-ablation and post-ablation, respectively. Abbreviations: r, pearson correlation coefficient; EV, estimated scar volume; MV, ground truth for scar volume; ESP, estimated scar percentage; MSP, ground truth for scar percentage.}
\label{fig:volume_sp_correlation} 
\end{figure}

\subsection{Analysis of Atrial Scars Quantification.}
\begin{figure}[!htp]
\begin{center}
\scalebox{.99}{
   \includegraphics[width=1\linewidth]{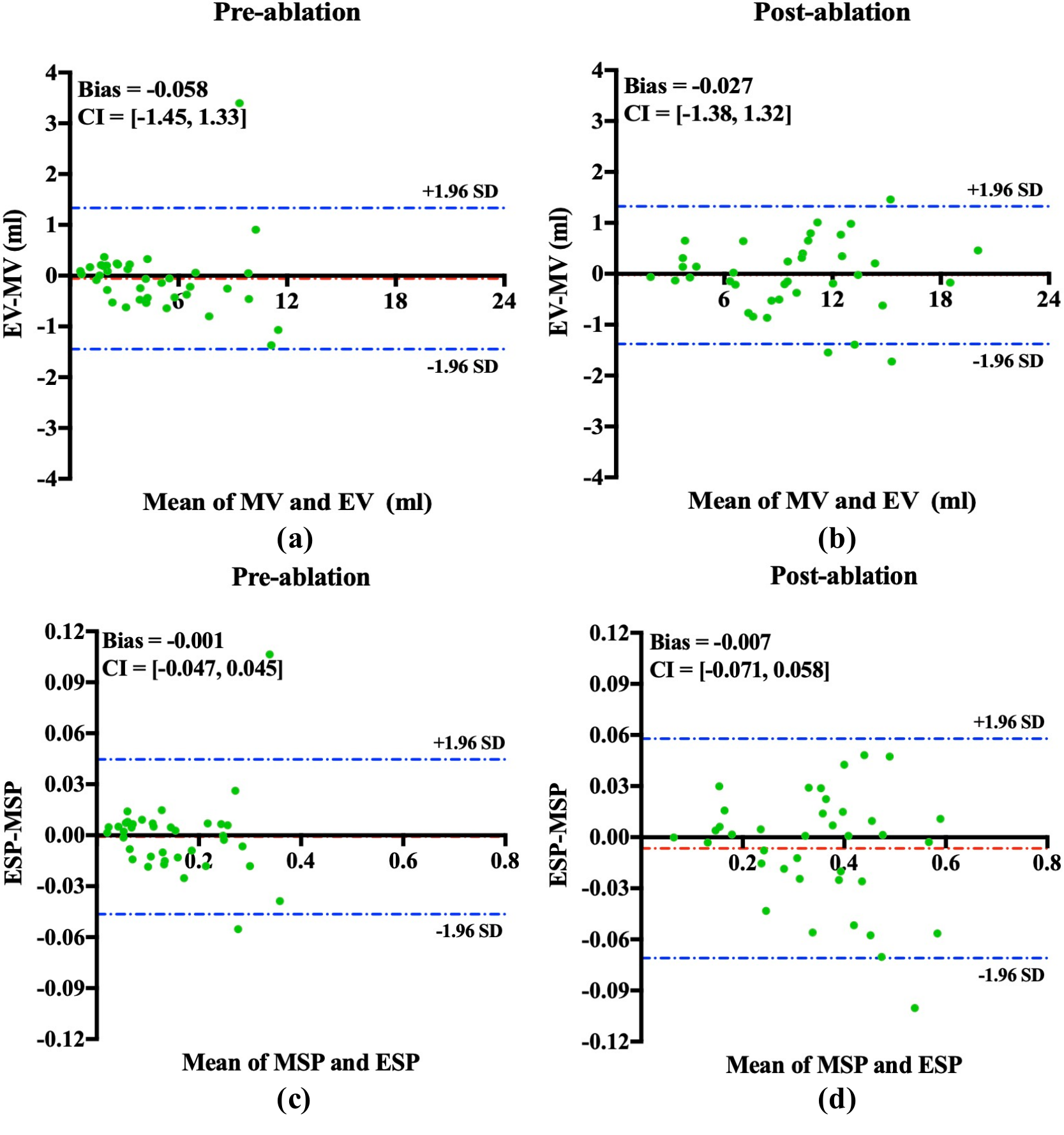}
   }
\end{center}
\caption{The agreement analysis based on Bland-Altman plots for JAS-GAN. (a) and (b) The high agreement between estimated scar volume and ground truth for pre-ablation and post-ablation, respectively. (c) and (d) The high agreement between estimated scar percentage and manual segmentation for pre-ablation and post-ablation, respectively. Abbreviations: EV, estimated scar volume; MV, ground truth for scar volume; ESP, estimated scar percentage; MSP, ground truth for scar percentage.}
\label{fig:agreement} 
\end{figure}

The quantification of the atrial scars is associated with scar percentage which is defined by the ratio of the scar volume to the LA wall volume. To measure the quantification results of atrial scars, we firstly reported scatter plots for the estimated scar volume and ground truth. As the linear regression results are shown in Fig. \ref{fig:volume_sp_correlation} (a) and (b), the Pearson correlation coefficients represented excellent correlation between the ground truth and our estimated results ($0.974$ for pre-ablation and $0.987$ for post-ablation). Besides, As the agreement results based on Bland-Altman plots are shown in Fig. \ref{fig:agreement} (a) and (b), our JAS-GAN was capable of estimating the scar volume with consistently low error. We then reported scatter plots for the estimated scar percentage and ground truth. As the linear regression results are shown in Fig. \ref{fig:volume_sp_correlation} (c) and (d), the Pearson correlation coefficients also showed the excellent correlation between the ground truth and our estimated results ($0.969$ for pre-ablation and $0.971$ for post-ablation). Furthermore, Fig. \ref{fig:agreement} (c) and (d) show the difference in calculated scar percentage against the scar percentage by manual segmentation. It is observed that the calculated scar percentage had a high agreement with manual delineation. These results indicated the ability of JAS-GAN for quantifying atrial scars.\

\begin{table*}
 \setlength{\abovecaptionskip}{0pt} 
 \setlength{\belowcaptionskip}{0pt} 
 \caption{\scriptsize{Compare the scar quantification and segmentation in LA on MICCAI 2018 Atrial Segmentation Challenge dataset. The results are presented in the form of mean $\pm$ standard deviation. Abbreviations: DSC, Dice Similarity Coefficient;  JI, Jaccard Index; ASD, Average Surface Distance; NMI, Normalized Mutual Information}}
 \centering
 \scalebox{.8}{
    \begin{tabular}{c|cccc|cccc}
    \addlinespace
    \toprule
     \multirow{2}*{Methods} & \multicolumn{4}{|c|}{LA and PVs} & \multicolumn{4}{c}{Atrial scars}\\ 
                           \cline{2-5}\cline{6-9}
                           & DSC & JI & ASD (mm) &NMI & DSC & JI & ASD (mm) &NMI \\
                           \hline
     2D U-Net & $0.898\pm0.034$ &$0.817 \pm 0.052$  &$3.38 \pm 4.53$ & $0.752\pm0.057$    &$0.526\pm0.118$ &$0.366\pm 0.109$  &$1.83 \pm 0.891$ & $0.396\pm0.112$ \cr 
     3D U-Net & $0.895\pm 0.032$ &$0.812 \pm 0.051$  &$3.81 \pm 3.89$ & $0.748\pm0.053$   &$0.508\pm 0.106$ &$0.347 \pm 0.096$  &$1.90 \pm 0.837$ & $0.383\pm0.093$\cr 
     Tversky loss & $0.904\pm 0.025$ &$0.826 \pm 0.041$ &$2.55 \pm 3.14$ & $0.764\pm0.046$    & $0.546\pm 0.093$ &$0.381 \pm 0.090$  &$1.88 \pm 0.838$ & $0.412\pm0.091$ \cr 
     Surface loss &$0.901\pm 0.039$ &$0.822 \pm 0.057$ & $2.73\pm 4.17$ & $0.758\pm0.063$     &$0.581\pm 0.114$ &$0.418 \pm 0.113$   & $1.51\pm 0.686$ & $0.448\pm0.112$ \cr
     MVTT &$0.902\pm 0.037$ &$0.823 \pm 0.057$ &$2.25\pm 1.39$ & $0.760\pm0.070$      &$0.613\pm 0.131$ &$0.454 \pm 0.132$ &$1.39\pm 1.03$ & $0.484\pm0.134$\cr
     JAS-GAN &$0.913\pm 0.027$ &$0.841 \pm 0.044$   & $2.24\pm 2.73$ & $0.782\pm0.049$   &$0.621\pm 0.110$ &$0.460 \pm 0.115$  & $1.24\pm 1.04$ & $0.489\pm0.116$\\ \midrule
    \end{tabular}
 }
 \label{table:challenge}
\end{table*}

\section{Discussion}
In this study, we have developed a JAS-GAN framework for the joint segmentations of unbalanced atrial targets of LA and atrial scars. The JAS-GAN framework consists of an adaptive attention cascade network and a joint discriminative network. In addition to the reported improvements by the ablation studies presented in Table \ref{table:ablation results}, extra analysis experiments were performed to further justify the rationale and the effectiveness of our used architecture illustrated in Section \uppercase\expandafter{\romannumeral4}. F and Section \uppercase\expandafter{\romannumeral4}. G. It is of note that the joint discriminative network only participates in the model training, thus will not increase the complexity of the final model in the testing phase or the practical applications.

Our proposed JAS-GAN framework is trained in an end-to-end manner based on the cascade connection with full supervision for the segmentations of unbalanced atrial targets, which provides an effective learning manner for the small atrial scars. We performed comprehensive experiments in the current study—comparing segmentation results of JAS-GAN with a two-phase segmentation for atrial scars with supervised learning (The automated segmented LA was used to define the ROI for the automated scar segmentation), and two two-phase methods with unsupervised learning (The manual segmented LA wall was used to define the ROI for the scar segmentation based on a classical method of the standard deviations thresholding (2SD) \cite{Karim2013Evaluation} and a state-of-the-art method of Ostu \cite{ravanelli2014novel}). As the results are shown in Table \uppercase\expandafter{\romannumeral1} and  Table \uppercase\expandafter{\romannumeral5}, the two-phase segmentation for atrial scars with supervised learning (RN+LA) improved the segmentation accuracy compared to the two-phase methods with unsupervised learning (2SD and Ostu), while our proposed JAS-GAN framework trained in the end-to-end manner with full supervision achieved the best segmentation accuracy. This is because that thresholding based 2SD and Ostu are the unsupervised methods, which are susceptible to noise. Because the atrial scars are very small, the noise hinders the accurate recognition of 2SD and Ostu for small atrial scars. Compared with unsupervised learning, deep learning based supervised learning can extract the high-level features to reduce the interference of noise for atrial scars identification. Furthermore, compared with the two-phase segmentation, the end-to-end learning manner is effective to relieve the problem that the inaccurate LA segmentation further leads to the inaccurate identification for atrial scars.

One limitation of our work is that our proposed method may not be applied directly to the external data if there are significant differences between our training data and the external testing data. This is a common issue while applying the deep learning algorithm to medical images in real clinical environment. Because the domain gaps widely exist between the training data and the external testing data if they come from different scanners or centres \cite{perone2019unsupervised,cheplygina2017transfer}. This problem may also be more severe for MRI based study because routinely used structural MRI (e.g., LGE MRI) are not quantitative acquisitions. Standardisation and normalisation of LGE MRI data can be problematic and an open question for research that is beyond the scope of our current study. To demonstrate that our proposed segmentation framework can be generalized to the data from different centres, we performed the experiments on the data from the MICCAI 2018 Atrial Segmentation Challenge with re-training model. The MICCAI 2018 Atrial Segmentation Challenge provided 100 scans with the labels of the LA wall and LA endocardium \cite{xiong2020global} (We directly combined the labels of LA wall and LA endocardium to obtain the label of LA epicardium. Then we automatically obtained the labels of atrial scars based on the protocol of Cardiac MRI Toolkit Slicer Extension from the National Alliance for Medical Image Computing). We randomly divided the data into a training set with 60 scans and a testing set with 40 scans for experiments.  Then we compared our proposed JAS-GAN with  the  widely  used  methods and the  state-of-the-art  methods (2D U-Net, 3D U-Net, Tversky loss, Surface loss and MVTT). As the experiment results are summarized in Table \ref{table:challenge}, our proposed JAS-GAN still achieved better segmentation accuracy for the LA and atrial scars, which indicated the application of our proposed JAS-GAN.

\section{Conclusion}
Automated and accurate segmentations of LA and atrial scars from LGE CMR images can provide great clinical significance for further quantifying atrial scars. In this study, we proposed a JAS-GAN model for the automated and accurate segmentations of the LA and atrial scars from LGE CMR images directly based on an adaptive attention cascade network and a joint discriminative network. The adaptive attention cascade network automatically captures the correlation of two segmentation tasks by building the relationship of LA and atrial scars. The joint discriminative network employs an adversarial regularization to force the  estimated  joint distribution of LA and atrial scars to match the real ones. The experimental results demonstrated that our proposed JAS-GAN enabled the accurate segmentations of the LA and atrial scars simultaneously. Therefore, our proposed JAS-GAN can provide an effective way in clinical practice to quantify the atrial scars for patients with AF.

\footnotesize
\bibliographystyle{ieeetr}
\bibliography{ref}

\begin{thebibliography}{10}

\bibitem{ravanelli2014novel}
D.~Ravanelli, E.~C. dal Piaz, M.~Centonze, G.~Casagranda, M.~Marini,
  M.~Del~Greco, {\em et~al.}, ``A novel skeleton based quantification and 3-d
  volumetric visualization of left atrium fibrosis using late gadolinium
  enhancement magnetic resonance imaging,'' {\em IEEE transactions on medical
  imaging}, vol.~33, no.~2, pp.~566--576, 2014.

\bibitem{Karim2013Evaluation}
R.~Karim, R.~J. Housden, M.~Balasubramaniam, C.~Zhong, D.~Perry, A.~Uddin, {\em
  et~al.}, ``Evaluation of current algorithms for segmentation of scar tissue
  from late gadolinium enhancement cardiovascular magnetic resonance of the
  left atrium: an open-access grand challenge,'' {\em Journal of Cardiovascular
  Magnetic Resonance}, vol.~15, no.~1, pp.~1--17, 2013.

\bibitem{vergara2011tailored}
G.~R. Vergara and N.~F. Marrouche, ``Tailored management of atrial fibrillation
  using a lge-mri based model: from the clinic to the electrophysiology
  laboratory,'' {\em Journal of cardiovascular electrophysiology}, vol.~22,
  no.~4, pp.~481--487, 2011.

\bibitem{siebermair2017assessment}
J.~Siebermair, E.~G. Kholmovski, and N.~Marrouche, ``Assessment of left atrial
  fibrosis by late gadolinium enhancement magnetic resonance imaging:
  methodology and clinical implications,'' {\em JACC: Clinical
  Electrophysiology}, vol.~3, no.~8, pp.~791--802, 2017.

\bibitem{khurram2016left}
I.~M. Khurram, M.~Habibi, I.~E. Gucuk, J.~Chrispin, E.~Yang, K.~Fukumoto, {\em
  et~al.}, ``Left atrial lge and arrhythmia recurrence following pulmonary vein
  isolation for paroxysmal and persistent af,'' {\em Jacc Cardiovascular
  Imaging}, vol.~9, no.~2, pp.~142--148, 2016.

\bibitem{xiong2020global}
Z.~Xiong, Q.~Xia, Z.~Hu, N.~Huang, C.~Bian, Y.~Zheng, S.~Vesal, N.~Ravikumar,
  A.~Maier, X.~Yang, {\em et~al.}, ``A global benchmark of algorithms for
  segmenting the left atrium from late gadolinium-enhanced cardiac magnetic
  resonance imaging,'' {\em Medical Image Analysis}, vol.~67, p.~101832, 2020.

\bibitem{isola2017image}
P.~Isola, J.-Y. Zhu, T.~Zhou, and A.~A. Efros, ``Image-to-image translation
  with conditional adversarial networks,'' in {\em Proceedings of the IEEE
  conference on computer vision and pattern recognition}, pp.~1125--1134, 2017.

\bibitem{hung2019adversarial}
W.~C. Hung, Y.~H. Tsai, Y.~T. Liou, Y.~Y. Lin, and M.~H. Yang, ``Adversarial
  learning for semi-supervised semantic segmentation,'' in {\em 29th British
  Machine Vision Conference, BMVC 2018}, 2019.

\bibitem{Yang2018Fully}
G.~Yang, X.~Zhuang, H.~Khan, S.~Haldar, E.~Nyktari, L.~Li, {\em et~al.},
  ``Fully automatic segmentation and objective assessment of atrial scars for
  longstanding persistent atrial fibrillation patients using late
  gadolinium‐enhanced mri,'' {\em Medical Physics}, 2018.

\bibitem{Perry2015Automatic}
D.~Perry, A.~Morris, N.~Burgon, C.~Mcgann, R.~Macleod, and J.~Cates,
  ``Automatic classification of scar tissue in late gadolinium enhancement
  cardiac mri for the assessment of left-atrial wall injury after
  radiofrequency ablation,'' {\em Proc Spie Int Soc Opt Eng}, vol.~8315,
  no.~8315, pp.~476--484, 2015.

\bibitem{karim2014method}
R.~Karim, A.~Arujuna, R.~J. Housden, J.~Gill, H.~Cliffe, K.~Matharu, {\em
  et~al.}, ``A method to standardize quantification of left atrial scar from
  delayed-enhancement mr images,'' {\em IEEE JTEHM}, 2014.

\bibitem{Mortazi2017CardiacNET}
A.~Mortazi, R.~Karim, K.~Rhode, J.~Burt, and U.~Bagci, ``Cardiacnet:
  Segmentation of left atrium and proximal pulmonary veins from mri using
  multi-view cnn,'' in {\em International Conference on Medical Image Computing
  and Computer-Assisted Intervention}, pp.~377--385, 2017.

\bibitem{Tobon2015Benchmark}
C.~Tobon-Gomez, A.~J. Geers, J.~Peters, J.~Weese, K.~Pinto, R.~Karim, {\em
  et~al.}, ``Benchmark for algorithms segmenting the left atrium from 3d ct and
  mri datasets,'' {\em IEEE Transactions on Medical Imaging}, vol.~34, no.~7,
  pp.~1460--1473, 2015.

\bibitem{XiongFully}
Z.~Xiong, V.~V. Fedorov, X.~Fu, E.~Cheng, R.~Macleod, and J.~Zhao, ``Fully
  automatic left atrium segmentation from late gadolinium enhanced magnetic
  resonance imaging using a dual fully convolutional neural network,'' {\em
  IEEE Transactions on Medical Imaging}, vol.~PP, no.~99, pp.~1--1, 2018.

\bibitem{chen2019discriminative}
J.~Chen, H.~Zhang, Y.~Zhang, S.~Zhao, R.~Mohiaddin, T.~Wong, {\em et~al.},
  ``Discriminative consistent domain generation for semi-supervised learning,''
  in {\em International Conference on Medical Image Computing and
  Computer-Assisted Intervention}, pp.~595--604, Springer, 2019.

\bibitem{yu2019uncertainty}
L.~Yu, S.~Wang, X.~Li, C.-W. Fu, and P.-A. Heng, ``Uncertainty-aware
  self-ensembling model for semi-supervised 3d left atrium segmentation,'' in
  {\em International Conference on Medical Image Computing and
  Computer-Assisted Intervention}, pp.~605--613, Springer, 2019.

\bibitem{zhuang2019evaluation}
X.~Zhuang, L.~Li, C.~Payer, D.~{\v{S}}tern, M.~Urschler, M.~P. Heinrich,
  J.~Oster, C.~Wang, {\"O}.~Smedby, C.~Bian, {\em et~al.}, ``Evaluation of
  algorithms for multi-modality whole heart segmentation: an open-access grand
  challenge,'' {\em Medical image analysis}, vol.~58, p.~101537, 2019.

\bibitem{li2017not}
X.~Li, Z.~Liu, P.~Luo, C.~Change~Loy, and X.~Tang, ``Not all pixels are equal:
  Difficulty-aware semantic segmentation via deep layer cascade,'' in {\em
  Proceedings of the IEEE conference on computer vision and pattern
  recognition}, pp.~3193--3202, 2017.

\bibitem{murthy2016deep}
V.~N. Murthy, V.~Singh, T.~Chen, R.~Manmatha, and D.~Comaniciu, ``Deep decision
  network for multi-class image classification,'' in {\em Proceedings of the
  IEEE conference on computer vision and pattern recognition}, pp.~2240--2248,
  2016.

\bibitem{cai2018cascade}
Z.~Cai and N.~Vasconcelos, ``Cascade r-cnn: Delving into high quality object
  detection,'' in {\em Proceedings of the IEEE conference on computer vision
  and pattern recognition}, pp.~6154--6162, 2018.

\bibitem{dai2016instance}
J.~Dai, K.~He, and J.~Sun, ``Instance-aware semantic segmentation via
  multi-task network cascades,'' in {\em Proceedings of the IEEE Conference on
  Computer Vision and Pattern Recognition}, pp.~3150--3158, 2016.

\bibitem{ouyang2017chained}
W.~Ouyang, K.~Wang, X.~Zhu, and X.~Wang, ``Chained cascade network for object
  detection,'' in {\em Proceedings of the IEEE International Conference on
  Computer Vision}, pp.~1938--1946, 2017.

\bibitem{lin2017cascaded}
L.~Di, G.~Chen, D.~Cohen-Or, P.-A. Heng, and H.~Huang, ``Cascaded feature
  network for semantic segmentation of rgb-d images,'' in {\em Proceedings of
  the IEEE International Conference on Computer Vision}, pp.~1311--1319, 2017.

\bibitem{chen2019hybrid}
K.~Chen, J.~Pang, J.~Wang, Y.~Xiong, X.~Li, S.~Sun, {\em et~al.}, ``Hybrid task
  cascade for instance segmentation,'' in {\em Proceedings of the IEEE
  Conference on Computer Vision and Pattern Recognition}, pp.~4974--4983, 2019.

\bibitem{Milletari2016V}
F.~Milletari, N.~Navab, and S.~A. Ahmadi, ``V-net: Fully convolutional neural
  networks for volumetric medical image segmentation,'' in {\em Fourth
  International Conference on 3d Vision}, 2016.

\bibitem{Ronneberger2015U}
O.~Ronneberger, P.~Fischer, and T.~Brox, ``U-net: Convolutional networks for
  biomedical image segmentation,'' in {\em MICCAI}, pp.~234--241, Springer,
  2015.

\bibitem{he2016deep}
K.~He, X.~Zhang, S.~Ren, and J.~Sun, ``Deep residual learning for image
  recognition,'' in {\em Proceedings of the IEEE conference on computer vision
  and pattern recognition}, pp.~770--778, 2016.

\bibitem{peters2009recurrence}
D.~C. Peters, J.~V. Wylie, T.~H. Hauser, R.~Nezafat, Y.~Han, J.~J. Woo, {\em
  et~al.}, ``Recurrence of atrial fibrillation correlates with the extent of
  post-procedural late gadolinium enhancement: a pilot study,'' {\em JACC:
  Cardiovascular Imaging}, vol.~2, no.~3, pp.~308--316, 2009.

\bibitem{haissaguerre1998spontaneous}
M.~Haissaguerre, P.~Ja{\"\i}s, D.~C. Shah, A.~Takahashi, M.~Hocini, G.~Quiniou,
  {\em et~al.}, ``Spontaneous initiation of atrial fibrillation by ectopic
  beats originating in the pulmonary veins,'' {\em New England Journal of
  Medicine}, vol.~339, no.~10, pp.~659--666, 1998.

\bibitem{keegan2015dynamic}
J.~Keegan, P.~D. Gatehouse, S.~Haldar, R.~Wage, S.~V. Babunarayan, and D.~N.
  Firmin, ``Dynamic inversion time for improved 3d late gadolinium enhancement
  imaging in patients with atrial fibrillation.,'' {\em Magnetic Resonance in
  Medicine}, vol.~73, no.~2, p.~646, 2015.

\bibitem{keegan2014improved}
J.~Keegan, P.~Jhooti, S.~V. Babu-Narayan, P.~Drivas, S.~Ernst, and D.~N.
  Firmin, ``Improved respiratory efficiency of 3d late gadolinium enhancement
  imaging using the continuously adaptive windowing strategy (claws),'' {\em
  Magnetic resonance in medicine}, vol.~71, no.~3, pp.~1064--1074, 2014.

\bibitem{keegan2014navigator}
J.~Keegan, P.~Drivas, and D.~N. Firmin, ``Navigator artifact reduction in
  three-dimensional late gadolinium enhancement imaging of the atria,'' {\em
  Magnetic resonance in medicine}, vol.~72, no.~3, pp.~779--785, 2014.

\bibitem{Salimans2016Improved}
T.~Salimans, I.~Goodfellow, W.~Zaremba, V.~Cheung, A.~Radford, and C.~Xi,
  ``Improved techniques for training gans,'' in {\em NIPS}, 2016.

\bibitem{kendall2018multi}
A.~Kendall, Y.~Gal, and R.~Cipolla, ``Multi-task learning using uncertainty to
  weigh losses for scene geometry and semantics,'' in {\em Proceedings of the
  IEEE conference on computer vision and pattern recognition}, pp.~7482--7491,
  2018.

\bibitem{dice1945measures}
L.~R. Dice, ``Measures of the amount of ecologic association between species,''
  {\em Ecology}, vol.~26, no.~3, pp.~297--302, 1945.

\bibitem{taha2015metrics}
A.~A. Taha and A.~Hanbury, ``Metrics for evaluating 3d medical image
  segmentation: analysis, selection, and tool,'' {\em BMC medical imaging},
  vol.~15, no.~1, p.~29, 2015.

\bibitem{miao2018image}
J.~Miao, T.-Z. Huang, X.~Zhou, Y.~Wang, and J.~Liu, ``Image segmentation based
  on an active contour model of partial image restoration with local cosine
  fitting energy,'' {\em Information Sciences}, vol.~447, pp.~52--71, 2018.

\bibitem{joskowicz2019inter}
L.~Joskowicz, D.~Cohen, N.~Caplan, and J.~Sosna, ``Inter-observer variability
  of manual contour delineation of structures in ct,'' {\em European
  radiology}, vol.~29, no.~3, pp.~1391--1399, 2019.

\bibitem{zamir2018taskonomy}
A.~R. Zamir, A.~Sax, W.~Shen, L.~J. Guibas, J.~Malik, and S.~Savarese,
  ``Taskonomy: Disentangling task transfer learning,'' in {\em Proceedings of
  the IEEE Conference on Computer Vision and Pattern Recognition},
  pp.~3712--3722, 2018.

\bibitem{cciccek20163d}
{\"O}.~{\c{C}}i{\c{c}}ek, A.~Abdulkadir, S.~S. Lienkamp, T.~Brox, and
  O.~Ronneberger, ``3d u-net: learning dense volumetric segmentation from
  sparse annotation,'' in {\em International conference on medical image
  computing and computer-assisted intervention}, pp.~424--432, Springer, 2016.

\bibitem{bui20173d}
T.~D. Bui, J.~Shin, and T.~Moon, ``3d densely convolutional networks for
  volumetric segmentation,'' {\em arXiv preprint arXiv:1709.03199}, 2017.

\bibitem{badrinarayanan2017segnet}
V.~Badrinarayanan, A.~Kendall, and R.~Cipolla, ``Segnet: A deep convolutional
  encoder-decoder architecture for image segmentation,'' {\em IEEE transactions
  on pattern analysis and machine intelligence}, vol.~39, no.~12,
  pp.~2481--2495, 2017.

\bibitem{yang2020simultaneous}
G.~Yang, J.~Chen, Z.~Gao, S.~Li, H.~Ni, E.~Angelini, T.~Wong, R.~Mohiaddin,
  E.~Nyktari, R.~Wage, {\em et~al.}, ``Simultaneous left atrium anatomy and
  scar segmentations via deep learning in multiview information with
  attention,'' {\em Future Generation Computer Systems}, vol.~107,
  pp.~215--228, 2020.

\bibitem{salehi2017tversky}
S.~S.~M. Salehi, D.~Erdogmus, and A.~Gholipour, ``Tversky loss function for
  image segmentation using 3d fully convolutional deep networks,'' in {\em
  International Workshop on Machine Learning in Medical Imaging}, pp.~379--387,
  Springer, 2017.

\bibitem{kervadec2019boundary}
H.~Kervadec, J.~Bouchtiba, C.~Desrosiers, E.~Granger, J.~Dolz, and I.~B. Ayed,
  ``Boundary loss for highly unbalanced segmentation,'' in {\em International
  conference on medical imaging with deep learning}, pp.~285--296, PMLR, 2019.

\bibitem{chen2018multi}
C.~Chen, W.~Bai, and D.~Rueckert, ``Multi-task learning for left atrial
  segmentation on ge-mri,'' in {\em International Workshop on Statistical
  Atlases and Computational Models of the Heart}, pp.~292--301, Springer, 2018.

\bibitem{perone2019unsupervised}
C.~S. Perone, P.~Ballester, R.~C. Barros, and J.~Cohen-Adad, ``Unsupervised
  domain adaptation for medical imaging segmentation with self-ensembling,''
  {\em NeuroImage}, vol.~194, pp.~1--11, 2019.

\bibitem{cheplygina2017transfer}
V.~Cheplygina, I.~P. Pena, J.~H. Pedersen, D.~A. Lynch, L.~S{\o}rensen, and
  M.~de~Bruijne, ``Transfer learning for multicenter classification of chronic
  obstructive pulmonary disease,'' {\em IEEE journal of biomedical and health
  informatics}, vol.~22, no.~5, pp.~1486--1496, 2017.

\end{thebibliography}

\appendix

\begin{figure}[!hbtp]
\begin{center}
\scalebox{.95}{
   \includegraphics[width=1\linewidth]{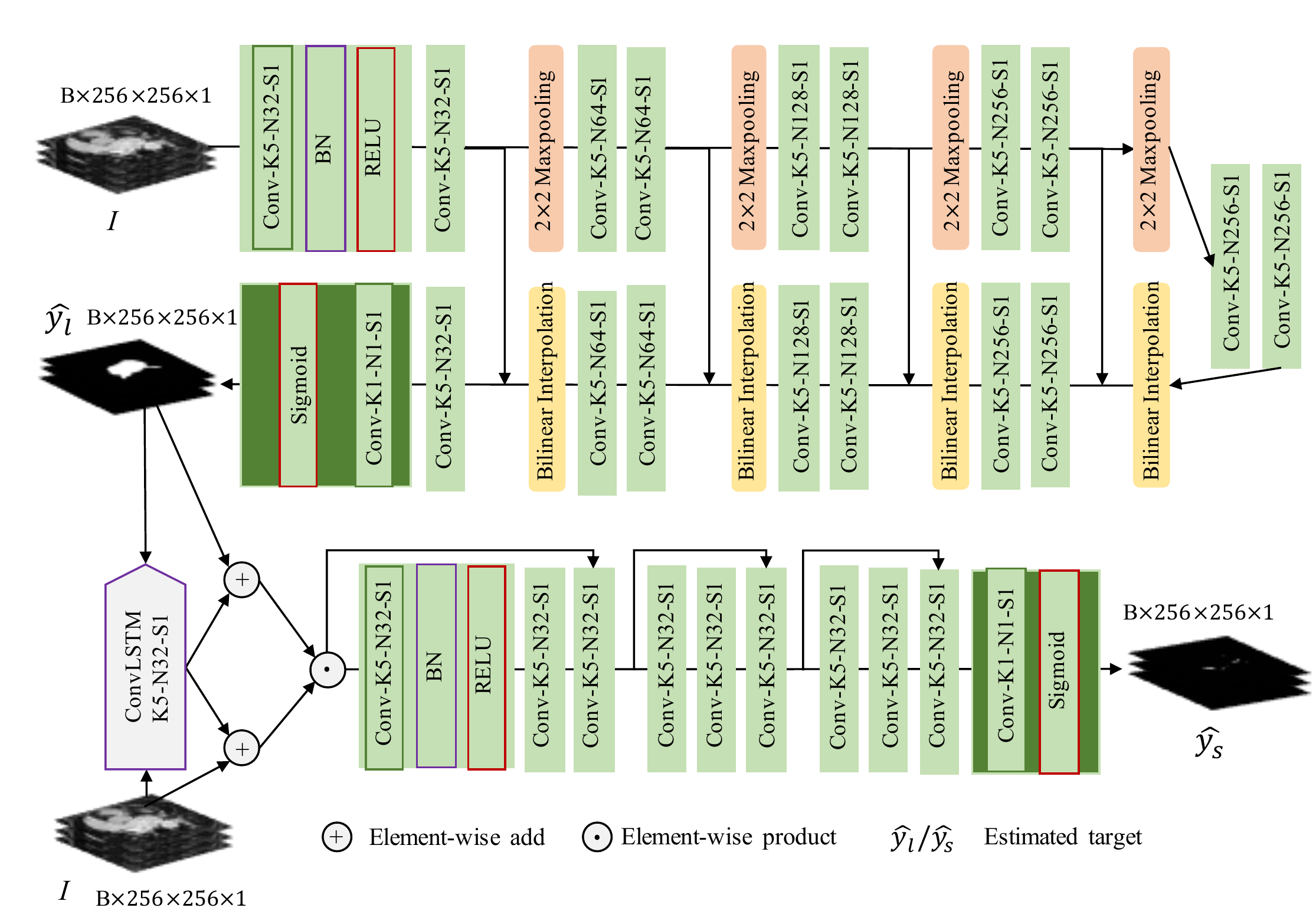}
   }
\end{center}
\caption{Architecture of the adaptive attention cascade network with corresponding kernel size (k), number of feature maps (n) and stride (s) indicated for each convolutional layer. Abbreviations: B, batchsize; BN, batch  normalization.}
\label{fig:RN} 
\end{figure}

\begin{figure}[!hbtp]
\begin{center}
\scalebox{.95}{
   \includegraphics[width=1\linewidth]{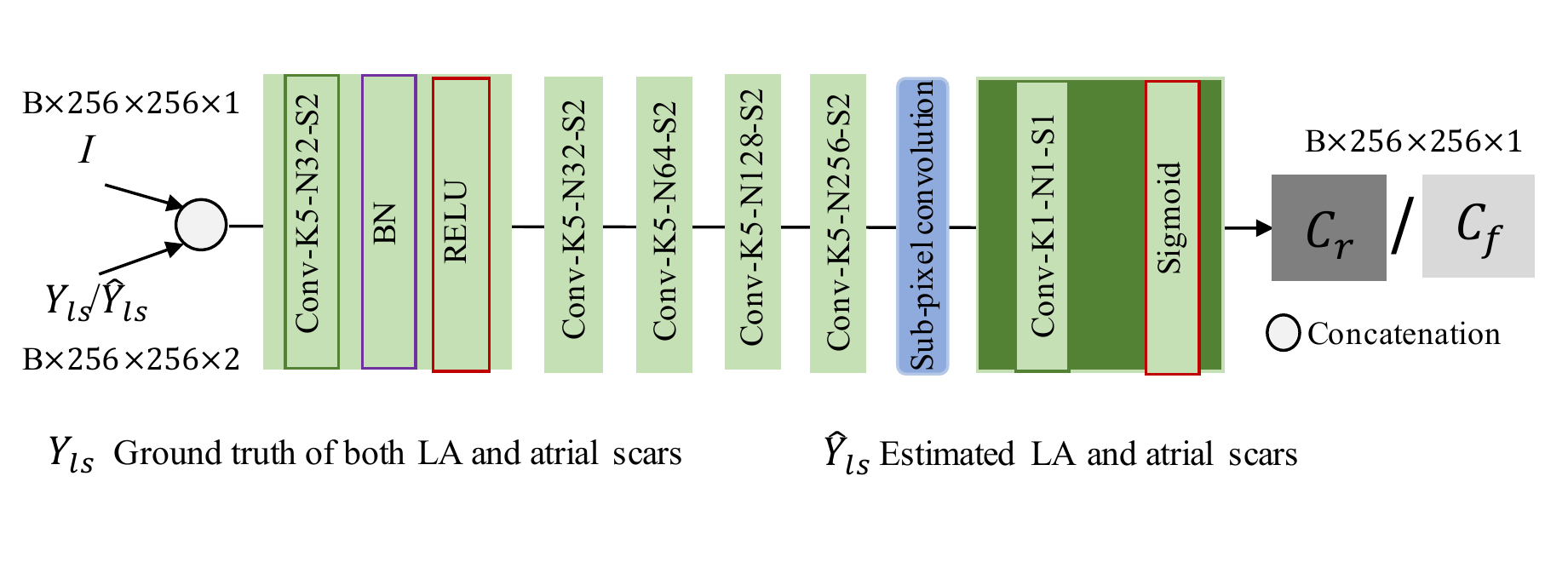}
   }
\end{center}
\caption{Architecture of the discriminative network with corresponding kernel size (k), number of feature maps (n) and stride (s) indicated for each convolutional layer. Abbreviations: B, batchsize; BN, batch  normalization.}
\label{fig:joint discriminator} 
\end{figure}



\end{document}